\documentclass[9pt, twocolumn, notitlepage]{article}

\usepackage{xspace}
\usepackage{xcolor}
\usepackage{graphicx}
\usepackage[noend]{algorithmic}
\usepackage{algorithm}
\usepackage[noend]{distribalgo}
\usepackage{subcaption}
\usepackage{multirow}
\usepackage{amssymb}
\usepackage{amsmath} 
\usepackage{authblk}
\usepackage[a4paper, margin=2cm]{geometry}

\newcommand{\lca}{$lca$\xspace}
\newcommand{\lcam}[1]{$#1.lca()$\xspace}
\newcommand{\msg}{\textsc{msg}\xspace}
\newcommand{\ack}{\textsc{ack}\xspace}
\newcommand{\notif}{\textsc{notif}\xspace}
\newcommand{\tpcc}{TPC-C\xspace}

\newcommand{\elia}[1]{{\color{black}#1}}

\newcommand{\la}{\leftarrow}
\newcommand{\flexcast}{FlexCast\xspace}

\newtheorem{lemma}{Lemma}
\newtheorem{proposition}{Proposition}

\newtheorem{definition}{Definition}

\begin{document}

\title{\flexcast: genuine overlay-based atomic multicast}

\author[1,4]{Eli\~{a} Batista}
\author[2]{Paulo Coelho}
\author[3]{Eduardo Alchieri}
\author[4]{Fernando Dotti}
\author[1]{Fernando Pedone}

\affil[1]{Universit\`{a} della Svizzera italiana, Lugano, Switzerland}
\affil[2]{Universidade Federal de Uberl\^{a}ndia, Uberl\^{a}ndia, Brazil}
\affil[3]{Universidade de Bras\'{i}lia, Bras\'{i}lia, Brazil}
\affil[4]{Pontif\'{i}cia Universidade Cat\'{o}lica do Rio Grande do Sul, Porto Alegre, Brazil}
\date{}                     
\setcounter{Maxaffil}{0}
\renewcommand\Affilfont{\itshape\small}







\maketitle

\begin{abstract}
Atomic multicast is a communication abstraction where messages are propagated to groups of processes with reliability and order guarantees.
Atomic multicast is at the core of strongly consistent storage and transactional systems.
This paper presents \flexcast, the first genuine overlay-based atomic multicast protocol.
Genuineness captures the essence of atomic multicast in that only the sender of a message and the message's destinations coordinate to order the message, leading to efficient protocols.
Overlay-based protocols restrict how process groups can communicate.
Limiting communication leads to simpler protocols and reduces the amount of information each process must keep about the rest of the system.
\flexcast implements genuine atomic multicast using a complete DAG overlay.
We experimentally evaluate \flexcast in a geographically distributed environment using gTPC-C, a variation of the TPC-C benchmark that takes into account geographical distribution and locality.
We show that, by exploiting genuineness and workload locality, \flexcast outperforms well-established atomic multicast protocols without the inherent communication overhead of state-of-the-art non-genuine multicast protocols.
\end{abstract}


\section{Introduction}

Atomic multicast is a communication abstraction that propagates messages to groups of processes with reliability and order guarantees.
Agreeing on the order of messages in the presence of failures is a notoriously difficult problem \cite{FLP85}.
Yet, message ordering is at the core of strongly consistent storage and transactional systems (e.g., \cite{corbett2012spanner,sciascia2012scalable,thomson2012calvin}).
Some systems implement strong consistency using an ad-hoc ordering protocol (e.g., \cite{granola-usenix12,corbett2012spanner}).
Atomic multicast encapsulates the logic for ordering messages and thereby reduces the complexity of designing fault-tolerant strongly consistent distributed systems.

In light of their important role, it is not surprising that many atomic multicast protocols have been proposed in the literature (e.g., \cite{defago04total,DGF00,rodrigues1998scalatom,fritzke1998amcast,schiper2008inherent}).
These protocols can be classified according to two criteria: (a) genuineness (or lack of) and (b) process connectivity.

\paragraph{Genuineness}
In a genuine atomic multicast protocol, only the message sender and destinations communicate to order a multicast message \cite{GS01b}. 
Some non-genuine atomic multicast protocols order messages using a fixed group of processes or involving all groups, regardless of the destination of the messages. 
In geographically distributed settings, a genuine atomic multicast protocol can better exploit locality than a non-genuine protocol since messages addressed to nearby groups do not introduce communication with remote groups.
Moreover, because a group only receives messages that are addressed to the group, in a genuine atomic multicast protocol groups do not incur communication overhead from relaying messages to the destinations.
This is important in geographically distributed environments where communication across wide-area links represents an important cost (e.g., Amazon Web Services).

\paragraph{Connectivity}
Most atomic multicast protocols assume that processes can communicate directly with one another. 
Alternatively, processes communicate following an \emph{overlay}, which determines which processes can exchange messages with which other processes.
Imposing limits on communication has advantages.
For example, overlays can represent the structure of administrative domains, simplify the design of protocols, and reduce the amount of information each process must keep about the rest of the system (e.g., key management in Byzantine fault tolerant protocols \cite{ByzCast}).

\vspace{2mm}
Combining genuineness and overlays is challenging.
Existing atomic multicast protocols focus on one aspect or the other but not both.
For example, all existing genuine atomic multicast protocols assume a fully connected overlay.
Hierarchical protocols, which structure communication between groups as a tree, are not genuine.
For example, in ByzCast \cite{ByzCast}, a multicast message is first sent to the lowest common ancestor of the message destinations, and then proceeds down the tree until it reaches all destinations.
ByzCast's logic is simple and processes in a group only need to keep information about their parent and children.
However, it is not genuine since a message addressed to the children of group $g$, but not to $g$, are first sent to $g$ and then propagated to $g$'s children, violating genuineness.

\begin{figure}[ht]
    \centering
    \includegraphics[width=0.8\columnwidth]{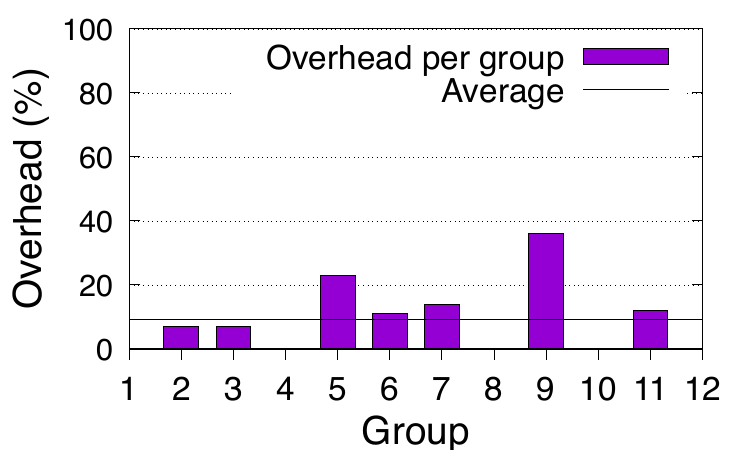}
    \caption{\small Communication overhead in a hierarchical protocol when executing the gTPC-C benchmark with tree $T_1$ and 90\% of locality (more details in Section \ref{sec:eval}); overhead, expressed as a percentage, is computed for each group as 1 minus the ratio between number of messages delivered and number of messages received by the group.}
    \label{fig:overhead}
\end{figure}

Figure \ref{fig:overhead} quantifies ByzCast's communication overhead, computed as one minus the ratio between the number of messages that a group delivers (i.e., messages addressed to the group) and the number of messages the group receives as part of communication imposed by the tree overlay, and expressed as a percentage.
On average, groups incur on almost 10\% of communication overhead.
Some groups, however, are more penalized than others, depending on their position in the tree.
In particular, about 23\% and 36\% of the communication of groups 5 and 9, respectively, is overhead.
This is in contrast to genuine atomic multicast protocols, which have no communication overhead.

\paragraph{Our contribution}
This paper proposes \flexcast, the first genuine overlay-based atomic multicast protocol. 
\flexcast assumes a complete directed acyclic graph (C-DAG) overlay.
Multicast messages are sent to the lowest common ancestor (\emph{lca}) of the message destinations.
The \emph{lca} then propagates the message to all other destinations in one communication step, without involving any groups that are not a message's destination.
\flexcast uses a sophisticated history-based protocol to order messages.
First, each process builds a history with all messages the process has delivered.
This history is propagated to other processes in the C-DAG, so that processes can ensure consistency (e.g., no two processes order two messages differently).
Simply following other processes' histories is not enough to ensure consistent order due to indirect dependencies.
Indirect dependencies happen for a few reasons.
For example, if process $x$ orders message $m_1$ before message $m_2$ and process $y$ orders $m_2$ before message $m_3$, then process $z$ must order $m_1$ before $m_3$ as a consequence of dependencies created by processes $x$ and $y$ involving $m_2$, a message not addressed to $z$.
\flexcast is well-suited to equip geographically replicated systems as it exploits locality.



We have implemented \flexcast and evaluated it in an emulated wide-area network that mimics Amazon's EC2.
To experimentally evaluate \flexcast, we propose gTPC-C, a variation of the well-known TPC-C benchmark that integrates geographical distribution.
In the original TPC-C benchmark, a transaction operates on items in a main warehouse and with a certain probability on items from additional warehouses as well.
gTPC-C models real-world wholesale supply systems in which transactions are directed to the customers' nearest warehouse and items not present in this warehouse are requested from the next closest warehouse and so on.
In gTPC-C, customers and warehouses are geographically distributed.
To account for locality, a customer's main warehouse is the closest one to the customer's location and multi-warehouse transactions have higher probability to involve warehouses located near the main warehouse.  
Our results show that, by exploiting locality, \flexcast can reduce latency by up to \elia{42\% to 46\%} when compared to state-of-the-art atomic multicast protocols in a geographically distributed environment.
Moreover, as a genuine atomic multicast protocol, \flexcast has no communication overhead.



The rest of the paper is structured as follows.
Section~\ref{sec:smodel} presents the system model and definitions used in the paper.
Section~\ref{sec:rwork} reports on related works.
Section~\ref{sec:idea} presents a detailed description of \flexcast, starting with a high level description of the protocol, then detailing the algorithms, and addressing practical concerns and fault tolerance.
Section~\ref{sec:eval} provides an experimental evaluation of \flexcast.
Section~\ref{sec:conclusion} concludes the paper.

\section{System model and definitions}
\label{sec:smodel}

This section presents our system model and recalls the definition of atomic multicast.

\subsection{System model}
\label{subsec:proccomm}

We consider a message-passing distributed system consisting of an unbounded set of client processes $C = \{ c_1, c_2, ... \}$ and a bounded set of server processes $S = \{ p_1, p_2, ... , p_n\}$. 
We define the set of server groups as $\Gamma = \lbrace G_A, G_B, ..., G_N \rbrace$, where for every $g \in \Gamma$, $g \subseteq S$.
Moreover, groups are non-empty and disjoint \cite{GS01b,gotsman2019white,schiper2008solving,ByzCast}. 
Processes are \emph{correct} if they never fail or \emph{faulty} otherwise.
In either case, processes do not experience arbitrary (i.e., Byzantine) behavior. 
We assume the system is partially synchronous \cite{DLS88}: it is initially asynchronous and eventually becomes synchronous. 
The time when the system becomes synchronous is called the Global Stabilization Time (GST), and it is unknown to the processes. 
Before GST, there are no bounds on communication and processing delays; after GST, such bounds exist but are unknown.

\subsection{Atomic multicast}
\label{subsec:atommcast}

Atomic multicast is a fundamental communication abstraction in reliable distributed systems. 
It encapsulates the complexity of reliably propagating and  ordering messages. 
With atomic multicast, a client can multicast messages to different groups with the guarantee that the destinations will deliver messages consistently.
In the following, we precisely capture these reliability and ordering guarantees.


A client atomically multicasts an application message $m$ to a set of groups by calling primitive $multicast(m)$, where $m.sender$ denotes the process that calls $multicast(m)$, $m.id$ is the message's unique identifier, and $m.dst$ is the groups $m$ is multicast to.
A server delivers message $m$ calling the primitive $deliver(m)$. 
If $|m.dst| = 1$ we say that $m$ is a \emph{local} message; if $|m.dst| > 1$ we say that $m$ is a \emph{global} message.

We define the relation $\prec$ on the set of messages server processes deliver as follows: 
$m \prec m'$ iff there exists a process that delivers $m$ before $m'$.
If $m \prec m'$ or $m' \prec m$, we say that there is a dependency between $m$ and $m'$.

Atomic multicast satisfies the following properties~\cite{hadzilacos1994modular}: 
\begin{itemize}
\item \textit{Validity}:~If a correct process $p$ multicasts a message $m$, then eventually all correct server processes $q \in g$, where $g \in m.\mathit{dst}$, deliver $m$.
\item \textit{Agreement}:~If a process $p$ delivers a message $m$, then eventually all correct server processes $q \in g$, where $g \in m.\mathit{dst}$, deliver $m$.
\item \textit{Integrity}:~For any process $p$ and any message $m$, $p$ delivers $m$ at most once, and only if $p \in g$, $g \in m.\mathit{dst}$, and $m$ was previously multicast.
\item \textit{Prefix order}:~For any two messages $m$ and $m'$ and any two server processes $p$ and $q$ such that $p \in g$, $q \in h$ and $\{ g, h \} \subseteq m.\mathit{dst} \cap m'.\mathit{dst}$, if $p$ delivers $m$ and $q$ delivers $m'$, then either $p$ delivers $m'$ before $m$ or $q$ delivers $m$ before $m'$.
\item \textit{Acyclic order}:~The relation $\prec$ is acyclic.
\end{itemize}

%

In a genuine atomic multicast protocol, only the sender and the destinations of a message coordinate to order the message. 
A genuine atomic multicast protocol does not depend on a fixed group of processes and does not involve processes unnecessarily. 
More precisely, a genuine atomic multicast algorithm should guarantee the following property \cite{GS01b}.

\begin{itemize}
    \item \emph{Minimality}: If a process $p$ sends or receives a message
    in run $R$, then some message $m$ is multicast in $R$, and $p$ is 
    $sender(m)$ or in a group in $m.dst$.
\end{itemize}

\section{Related work}
\label{sec:rwork}

An early atomic multicast protocol is attributed to D. Skeen \cite{Birman}.
In this protocol, a multicast message $m$ is first propagated to $m$'s destinations.
Upon receiving the message, a destination assigns the message a local timestamp and sends the local timestamp to the other message destinations.
When a destination has received timestamp from all message destinations, it computes the message's final timestamp as the maximum among all of the message's local timestamps.
Destinations deliver messages in order of their final timestamp.
This protocol is genuine but does not tolerate failures.

Several atomic multicast protocols extend Skeen's ordering technique to tolerate failures \cite{FastCast17}, \cite{fritzke1998amcast}, \cite{gotsman2019white}, \cite{Le2021RamCastRA}, \cite{rodrigues1998scalatom}. 
In all these protocols, the idea is to implement destinations as groups of processes.
Thus, messages are addressed to one or more process groups, instead of a set of processes, as in the original protocol.
Although some processes in a group may fail, each group acts as a reliable entity, whose logic is replicated within the group using state machine replication \cite{Sch90}.
Recent protocols aim at reducing the cost of replication within groups while keeping Skeen's original idea of assigning timestamps to messages and delivering messages in timestamp order.
FastCast \cite{FastCast17} improves performance by optimistically executing parts of the replication logic within a group in parallel.
WhiteBox\cite{gotsman2019white} atomic multicast uses the leader-follower approach to replicate processes within groups.
RamCast \cite{Le2021RamCastRA} relies on distributed shared memory (RDMA) to reduce latency.
Since in all these protocols processes communicate directly with one another, we refer to them as \emph{distributed} atomic multicast protocols (see Table~\ref{tbl:amcasts}).

\begin{table}[h]
\footnotesize
\centering
\begin{tabular}{|l|l|l|} \hline
{\bf Class}		& {\bf Type}	& {\bf Examples} \\ \hline
Distributed	& genuine		& \cite{Birman, FastCast17, fritzke1998amcast, DGF00, gotsman2019white, Le2021RamCastRA, rodrigues1998scalatom} \\ \hline
Hierarchical	& non-genuine	& \cite{ByzCast, 37965, herlihy2006arrow} \\ \hline
C-DAG overlay		& genuine		& FlexCast (this paper)\\ \hline
\end{tabular}
\caption{Different classes of atomic multicast protocols.}
\label{tbl:amcasts}
\end{table}%

In~\cite{DGF00}, a genuine distributed atomic multicast protocol that does not rely on exchanging of timestamps to order messages is proposed.
The protocol assigns a total order to groups and relays messages sequentially through their destination groups following this order.
A multicast message $m$ is initially sent to the lowest group in $m.dst$ according to the total order.
When the group receives $m$, it uses consensus to order and deliver $m$ inside the group, then $m$ is forwarded to the next group in $m.dst$, according to the total order of groups.
A group that delivers $m$ can only order the next message once it knows $m$ is ordered in all groups in 
$m.dst$, which is after it receives an \textsc{end} message from the last group in $m.dst$.   
Although the dissemination of the message follows an order, the \textsc{end} message returns
to each group involved and therefore the protocol is a distributed atomic multicast protocol. 
Besides needing $n+1$ steps to deliver a message, where $n$ is the number of destinations of 
the message, since groups remain locked until the \textsc{end} message arrives, this protocol
is affected by the convoy effect \cite{7600156}.

Some protocols restrict process communication by means of a tree overlay that determines how groups can communicate (e.g., \cite{ByzCast,37965}).
To order a message $m$ using a tree, $m$ is first sent to the lowest common ancestor group among those in $m.dst$, in the worst case the root of the overlay tree. 
Then, $m$ is successively ordered by the lower groups in the tree until it reaches all groups in $m.dst$. 
An important invariant is that lower groups in the tree preserve the order induced by higher groups.
Although simple, this protocol is not genuine since a message may need to be ordered by a group that is not in the destination set of the message.
While the tree-based protocol proposed in \cite{37965} does not tolerate failures, ByzCast \cite{ByzCast} can withstand Byzantine failures.

\begin{figure*}[h]
	\centering
    \includegraphics[width=1.2\columnwidth]{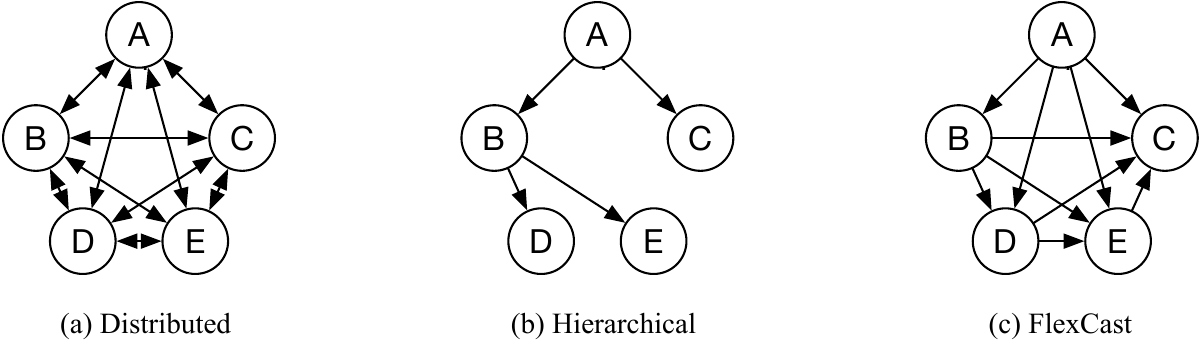}
    \caption{\small Three communication patterns used in atomic multicast protocols involving groups $A, B, ..., E$: 
    (a)~distributed,
    (b)~hierarchical, and 
    (c)~FlexCast, the approach presented in this paper.
    In the graphs, directed edge $g \rightarrow h$ means that group $g$ can send messages to group $h$, and $h$ can receive messages from $g$ but not send messages to $g$.}
    \label{fig:overlays}
\end{figure*}

The Arrow \cite{herlihy2006arrow} protocol is a non-fault tolerant tree-based protocol that targets open groups.
It emerges from the combination of a reliable multicast protocol with a distributed swap protocol.
Arrow assumes a graph $G$ and a spanning tree $T$ on $G$.
Initially, each node $v$ in $T$ has $link(v)$ that is its neighbour in $T$ or
itself if $v$ is a sink (initially only the root of $T$).  
To multicast $m$ a node $v$ sends a message through $link(v)$, which
is forwarded to the root of the tree.   By definition,
the root has sent the last message before $m$.   As the message
is forwarded, edges change direction and $v$ becomes the new root
(that has sent the last message, which now is $m$).
Although genuine, this procedure may result in swap
messages traversing the diameter of $T$ and only then a multicast, using an
underlying reliable multicast, is issued.

Restricting communication as in a tree may lead to simpler atomic multicast algorithms.
Moreover, if communication needs to be authenticated, as in Byzantine fault-tolerant protocols, a tree overlay requires fewer keys to be maintained and exchanged between processes than a distributed fully connected protocol.
Finally, a fully connected protocol is a reasonable assumption in systems that run within the same administrative domain (e.g., Google's Spanner [14]).
In other contexts (e.g., decentralized systems), however, multiple entities from different administrative domains collaborate but do not wish to establish connections with all other domains.
Hereafter, we refer to protocols based on a tree as \emph{hierarchical} atomic multicast protocols.

Figure~\ref{fig:overlays} shows three cases of interest.
All genuine atomic multicast algorithms we are aware of are distributed (Figure~\ref{fig:overlays}~(a)).
A tree (Figure~\ref{fig:overlays}~(b)) is the minimum connectivity needed by any atomic multicast protocol to support an arbitrary workload (i.e., messages can be multicast to any set of groups), as removing one edge from the tree results in a partitioned graph.
Hierarchical protocols, however, are not genuine.
For example, in Figure~\ref{fig:overlays}~(b), a message multicast to groups $B$ and $C$ will first be ordered at $A$, and then propagated and ordered by  $B$ and $C$.
This paper proposes the first overlay-based genuine atomic multicast protocol.

\section{Genuine overlay-based atomic multicast}
\label{sec:idea}

In this section, we present \flexcast's basic idea and detailed algorithm, and conclude with practical considerations and a discussion on fault tolerance.
\flexcast's correctness is presented in the appendix of this paper.

\subsection{General idea}
\label{subsec:basic}

Groups in \flexcast are structured as a complete directed acyclic graph (C-DAG), as the example in Figure~\ref{fig:overlays}~(c). 
We assume there is a total order among groups. Each group is assigned a unique rank in $0..(n-1)$, 
where $n$ is the number of groups.   The C-DAG topology is such that there is a directed edge from 
each group with rank $i$ to each group with rank $j$ if $i < j$.
In this graph, $i$'s \emph{ancestors} have lower rank than $i$ and $i$'s \emph{descendants} have higher rank than $i$.\footnote{We use the terms ``lower" and ``higher" groups to denote relative positions of groups in this rank, and ``lowest" and ``highest" group of a subset of groups, also referring to this rank.  ``Ancestors" of a group $g$ denote the set of groups lower than $g$, while ``descendants" respectively higher.}   Figure~\ref{fig:overlays}~(c) shows a C-DAG with nodes ordered from lowest to highest as: A, B, D, E, C.

A client atomically multicasts a message $m$ by sending $m$ to $m$'s lowest common ancestor (\lca).
The \lca of a multicast message is the group with the lowest rank among the destinations of the message.
At its \lca, $m$ is directly delivered and propagated to $m$'s other destination groups 
(by definition the \lca has direct edges with each other destination group in $m.dst$).
Similarly to a tree-base atomic multicast, in a C-DAG, a group must respect the dependencies created by its ancestors and propagate dependencies to its descendants.
In a C-DAG, however, a group may have multiple ancestors and dependencies can be created by any of them.
An important challenge is to ensure that dependencies are properly communicated down the C-DAG without violating the minimality property of genuine atomic multicast.
\flexcast uses three strategies to accomplish this, as explained next.

\vspace{1mm}
{\it Strategy (a):  } First, every group keeps track of a \emph{history}, a graph where messages are vertexes and their relative order are edges.
A vertex contains a message's id and destinations.   
Messages delivered at a group are recorded in its history and build a total order within the graph.  
When a group propagates a message to another one, its history  is included.   The destination group
extends its history with the histories that it receives from other groups and messages it delivers.
The history then becomes a graph. More specifically, since ordering is respected (discussed next), the history is a DAG.
Destination groups use the history to ensure that messages are delivered consistently across the system.

\begin{figure*}[ht]
	\centering
    \includegraphics[width=2\columnwidth]{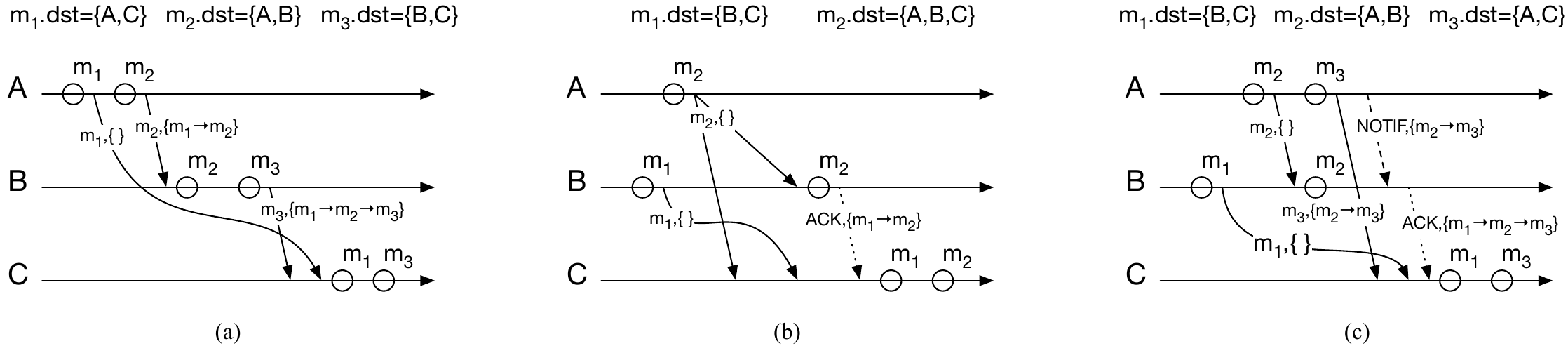}
    \caption{\small Executions of \flexcast illustrating the use of (a) histories, (b) \ack messages, and (c) \notif messages in an overlay where $A \rightarrow B, A \rightarrow C$ and $B \rightarrow C$. (Legend: a full arrow is the propagation of an application message, a circle is the delivery of a message, a dotted arrow is an \ack message, and a dashed arrow is a \notif message). }
    \label{fig:threecases}
\end{figure*}

To understand the need for exchanging histories, consider the scenario depicted in Figure \ref{fig:threecases}~(a), where group $A$ is the \lca of messages $m_1$ (multicast to $A$ and $C$) and $m_2$ (multicast to $A$ and $B$), and group $B$ is the \lca of $m_3$ (multicast to $B$ and $C$).
Since $A$ delivers $m_1$ before $m_2$ (i.e., $m_1 \prec m_2$) and $B$ delivers $m_2$ before $m_3$ (i.e., $m_2 \prec m_3$), $C$ must deliver $m_1$ before $m_3$ to avoid a cycle among delivered messages.
But $C$ receives $m_3$ from $B$ before it receives $m_1$ from $A$.
By receiving $B$'s history, $C$ knows that it should deliver $m_1$ and then $m_3$ to avoid cycles.

Unfortunately, including histories in forwarded messages is not enough to avoid cycles. 
Intuitively, this happens because not all dependencies are captured in the communication of application messages between groups.
There are two cases to consider, depending on whether the group that creates the dependency is aware that it must propagate the dependency to its descendants or not.

\vspace{1mm}
{\it Strategy (b):  }To motivate the case where a group is aware that it should send dependencies to its descendants, consider the execution in Figure \ref{fig:threecases}~(b).
In this case, $B$ delivers $m_1$ before $m_2$, and $C$ receives $m_2$ from $A$ (with an empty history) and then $m_1$ from $B$ (with an empty history since $B$ did not know about $m_2$ when it sent $m_1$ to $C$).
Yet, $C$ must deliver $m_1$ before $m_2$.
\flexcast ensures proper order in such cases as follows.
If group $g$ and its descendant $h$ are in the destination of a message $m$ and $g$ is not $m$'s \lca, then 
$g$ sends an \ack message to $h$ with $g$'s history.
Conversely, if $h$ receives a message $m$ and $h$ has an ancestor that is in $m$'s destination, but is not $m$'s \lca, $h$ waits for $g$'s 
\ack message.

\vspace{1mm}
{\it Strategy (c):  }To motivate the case where a group is not aware that it should send dependencies to its descendants, consider the execution in Figure \ref{fig:threecases}~(c).
In this case, group $A$ sends $m_3$ and its history (i.e., $m_2$ precedes $m_3$) to $C$, and $B$ sends $m_1$ and an empty history to $C$ (i.e., because the dependency between $m_1$ and $m_2$ happens in $B$ after $B$ communicates with $C$).
$B$ does not send $C$ the information that $m_1$ precedes $m_2$ since $m_2$ is not addressed to $C$.
Yet, $C$ must deliver $m_1$ before $m_3$.
To handle this case,
when a group determines that a descendant $d$ must forward its history down the C-DAG, it sends a \notif message to $d$ so that $d$ can communicate its dependencies to other groups.

More precisely, when a group $g$, the \lca of a message (or another destination in $m.dst$) is about to forward message $m$ (respectively, an \ack message regarding $m$) and there is a group $h$ such that:
(i) $h$ is not in $m.dst$; 
(ii) $h$ is a descendant of $g$ and an ancestor of group $r$ in $m.dst$; and 
(iii) there is a message in $g$'s history addressed to $h$, 
then $g$ sends a \notif message regarding $m$ to $h$.
If group $h$ receives a \notif message regarding $m$, it sends \ack messages to all its descendants $k \in m.dst$.
Moreover, inductively, if there is a message $h'$ in $h$'s history with the same restrictions above, $h$ notifies $h'$.
This induction naturally finishes since there is a total order on groups.


\subsubsection{Why it is genuine}
To argue that \flexcast is genuine, first notice the following aspects discussed about \textit{Strategies (a)} and \textit{(b)}:

\begin{itemize}
\item when $m$ is multicast, it enters the overlay at \lcam{m} (see Algorithm \ref{algo:structs}), which is by definition a destination of $m$;
\item \lcam{m} propagates $m$ to its further destinations in $m.dst$; and
\item each destination $d$ (other than \lcam{m}) sends \ack messages to groups in $m.dst$ higher than $d$.
\end{itemize}
From the above, it follows that the communication described involves exclusively groups in $m.dst$.

Now, consider the \textit{Strategy (c)} and notice that:
\begin{itemize}
\item a group $g \in m.dst$ can send a \notif message to a group $h \notin m.dst$ provided that
 $g$ previously sent a message to $h$, 
i.e. some message was multicast to $h$ in run $R$; and
\item inductively, $h$ notifies $h'$ only if some message was multicast from $h$ to $h'$ in run $R$.
\end{itemize}
From the above, it follows that groups not in $m.dst$ exchange messages only if they communicated
in run $R$, keeping minimality (see definition in Section \ref{subsec:atommcast}).

\subsection{Detailed protocol}
\label{subsec:algospec}



Algorithm \ref{algo:structs} presents the basic data structures used in \flexcast. 
Each group knows the C-DAG topology and has a communication channel to each descendant group (i.e., a FIFO reliable point-to-point link).
As a consequence, each process has an input queue for each input channel from ancestor groups (line \ref{algstruct:queues}).
Each queue contains not-yet-delivered messages sent by the respective ancestors.

A message has a unique $id$ (line \ref{algstruct:mid}), a set of destination groups (line \ref{algstruct:dst}), and an arbitrary payload (line \ref{algstruct:pload}), provided by the application.
The protocol stores pending messages along with a set of respective \ack messages (line \ref{algstruct:acks}) and a set of notified groups (line \ref{algstruct:nl}), both detailed later. 
Function $m.lca()$ (line \ref{algstruct:lca}) returns the lowest group in $m.dst$.

A group $g$ has the history it learns from each of its ancestors and the messages it delivers (line \ref{algstruct:hst}).
The set of messages delivered in $g$ is a subset of messages in the history (line \ref{algstruct:deliveredInG}).
The history builds a DAG with dependencies in $hst.D$.
As notification messages may not be immediately delivered according to criteria to be detailed later,
a group also has a set of pending notification messages (line \ref{algstruct:pn}).

When group $g$ communicates with a descendent group $h$, $g$
informs only the difference in $g$'s history with respect to the last message $g$ sent to $h$.   
Therefore, for each descendent $h$, $g$ keeps track of what part 
of its history it has already sent to $h$ (line \ref{algstruct:hsth}).     
  
\begin{algorithm}
    \caption{Types and data structures, for each group g}
    \label{algo:structs}
    \begin{distribalgo}[1]
    \footnotesize 
    
    

    \INDENT{\textbf{Type $Message$}: every message $m$ has:}
        \STATE $m.id$ \label{algstruct:mid} \COMMENT {$m$'s global unique id}
        \STATE $m.dst$ \label{algstruct:dst} \COMMENT {$m$'s destinations, a subset of groups}
        \STATE $m.payload$ \label{algstruct:pload} \COMMENT {provided by the application}
        \STATE $m.acks \la \varnothing$ \label{algstruct:acks} \COMMENT {a set of received acks}
        \STATE $m.notifList \la \varnothing$ \label{algstruct:nl} \COMMENT {a set of notified groups}
        \STATE $m.lca():func$ \label{algstruct:lca} \COMMENT {returns the lca in $m.dst$}
    \ENDINDENT
    
    \vspace{1mm}
        \INDENT[a history is ]{\textbf{Type (history) $H$}:}
        \STATE $H = (M, D, lastDlvd)$                 \COMMENT{messages, dependencies, last one}
        \STATE $M : $ set of $Message$ \label{algstruct:M}     \COMMENT {a pair $(m_1,m_2) \in D$ means ...}
        \STATE $D : M \times M$ \label{algstruct:dep} \COMMENT {$m_1$ ordered before $m_2$: $m_2$ depends of $m_1$}
        \STATE $lastDlvd: M \cup \{ \bot \}$        \COMMENT{the last message delivered}
      \ENDINDENT
    
    \vspace{1mm}
    \INDENT{\textbf{Group $g$ variables:}}	
        \STATE $queues \la [\varnothing, ..., \varnothing]$ \label{algstruct:queues} \COMMENT{an empty queue per ancestor}   
        \STATE $hst \la H(\varnothing, \varnothing, \bot)$ \label{algstruct:hst} \COMMENT {the initial history of group g}
        \STATE $deliveredInG \subseteq hst.M$ \label{algstruct:deliveredInG} \COMMENT {the messages in hst delivered in g}
        \STATE $pendNotif \la \varnothing$ \label{algstruct:pn} \COMMENT {a set of pending notifications}
       \STATE $\forall ~ h$ higher than $g, hst(h) \la H(\varnothing, \varnothing, \bot)$ \label{algstruct:hsth} \COMMENT {the history of $g$ \\    
       \hfill informed to each $h$ so far}
    \ENDINDENT
    
    \end{distribalgo}
\end{algorithm}

To atomic multicast message $m$, a client sends $m$ to $m.lca()$.
Algorithm~\ref{algo:events} presents the events triggered at a group when receiving each one of the three types of messages in our protocol: (i) \msg is a client message; (ii) \ack is an acknowledge message; and (iii) \notif is a notification message.
Algorithm \ref{algo:funcs} presents the core functions used in Algorithm~\ref{algo:events}.

\begin{algorithm}
    \caption{Events, for each group g}
    \label{algo:events}
    \begin{distribalgo}[1]
    \footnotesize 
    
    \vspace{1mm}

    \INDENT
    {\textbf{upon} receiving $[\msg, m, history] \land  g = m.lca()$} \label{algo:events:msglca}
             \STATE $a$-$deliver(m)$ \label{algo:events:deliver}
     \ENDINDENT
       \vspace{1mm}
    
  \INDENT
    {\textbf{upon} receiving $[\msg, m, history] \land g \neq m.lca()$} \label{algo:events:msg}
            \STATE $update$-$hst(history)$  \label{algo:events:updt}
            \STATE $queues[m.lca()].enqueue(m)$  \label{algo:events:enq}
            \STATE $reprocess$-$queues()$  \label{algo:events:reproc}
            
    \ENDINDENT
    
    \vspace{1mm}

    \INDENT
    {\textbf{upon} receiving $ [\ack, m, history]$ from ancestor $a$}
    \label{algo:events:ack}
            \STATE $update$-$hst(history)$ \label{algo:events:updtack}
            \STATE $queues[m.lca()].get(m.id).acks.add([\ack$ from $a])$ \label{algo:events:addack}
            \STATE $queues[m.lca()].get(m.id).$\textit{notifList}.merge(\textit{m.notifList}) \label{algo:events:mergeNotifList}
            \STATE $reprocess$-$queues()$ \label{algo:events:repack}
    \ENDINDENT
    
    \vspace{1mm}
    \INDENT
    {\textbf{upon} receiving $ [\notif, m, history]$ }
    \label{algo:events:notif}
        \STATE $update$-$hst(history)$ \label{algo:events:updtnotif}
        \STATE $deps \leftarrow open$-$dependencies()$
        \IF{$deps \neq \varnothing$} \label{algo:events:incDeps}
            \STATE $pendNotif.add([\notif, m, deps])$ \label{algo:events:addpend}
        \ELSE
            \STATE $send$-$descendants(m, \ack)$ \label{algo:events:sendack}
        \ENDIF
    \ENDINDENT

    \end{distribalgo}

\end{algorithm}

In \flexcast, the $lca$ delivers a multicast message as soon as it receives the message.
In doing so, the $lca$ imposes its delivery order on all its descendant groups through information disseminated in histories and auxiliary messages.
Upon receiving a multicast message $m$, if $g$ is the \lca (line \ref{algo:events:msglca}), $g$ can deliver $m$ immediately (line \ref{algo:events:deliver}). 

When non \lca groups receive a \msg (line \ref{algo:events:msg}) first they update their local history with the history received together with $m$ (line \ref{algo:events:updt}), enqueue $m$ in the corresponding ancestor's queue (line \ref{algo:events:enq}), and reprocess all ancestors' queues (line \ref{algo:events:reproc}), since this message may carry the information needed to deliver other messages. 

When receiving an \ack message (line \ref{algo:events:ack}), $g$ updates its local history (line \ref{algo:events:updtack}), and associates the \ack to the multicast message $m$ in the \lca's queue that originated the \ack (line \ref{algo:events:addack}). 
Since an \ack may identify further groups to be notified,
the message's list of notified groups is updated accordingly (line \ref{algo:events:mergeNotifList}).
Group $g$ then reprocesses all queues (line \ref{algo:events:repack}).

When receiving a \notif message (line \ref{algo:events:notif}), $g$ updates its local history (line \ref{algo:events:updtnotif}), sends the necessary \ack messages (line \ref{algo:events:sendack}), and possibly sends notification messages to its descendants as well, as detailed later. 
However, if the local history contains a message $m'$ addressed to $g$ that was not delivered yet,
then $g$ waits until it delivers $m'$ before sending the \ack messages, and appends the \notif in the pending notifications set 
(line \ref{algo:events:addpend}), avoiding propagating incomplete dependencies.

In Algorithm \ref{algo:funcs}, when $g$ delivers a message, it adds the message to its history (line \ref{algo:funcs:hstAdd}).
The total order of delivered messages is built having the new message depend on the last
message delivered (lines \ref{algo:funcs:buildTO1} and \ref{algo:funcs:buildTO2}).

\begin{algorithm}[H]
    \caption{Main logic, for each group g}
    \label{algo:funcs}
    \begin{distribalgo}[1]
    \footnotesize

       \vspace{1mm}
    \INDENT[ancestor's history $ah$] {\textbf{update-hst} $(ah: H)$} \label{algo:funcs:update-hst}
        \STATE{$hst.M \la hst.M \cup ah.M$}          \COMMENT{messages and dependencies are}
        \STATE{$hst.D \la hst.M \cup ah.D$}           \COMMENT{intergated to the group's $hst$}
    \ENDINDENT

    \vspace{1mm}
    \INDENT {\textbf{hst-add} $(m: Message)$} \label{algo:funcs:hstAdd}
        \STATE{$hst.M \la hst.M \cup \{ m \}$}                        \COMMENT{add $m$, if not yet in hst}
        \STATE{$hst.D \la hst.D \cup \{ (hst.lastDlvd, m) \}$}  \COMMENT{build total order in}              \label{algo:funcs:buildTO1}
         \STATE{$hst.lastDlvd \la m $}                                    \COMMENT{msgs delivered at this group}  \label{algo:funcs:buildTO2}
         \STATE{$deliverdInG \la deliverdInG \cup \{ m \}$}    \label{algo:funcs:deliverInG}
     \ENDINDENT  
                
       \vspace{1mm}
     \INDENT {\textbf{open-dependencies} $()$: set of Messages} \label{algo:funcs:hasIncompleteDeps}
        \STATE{\textbf{return} \{$\forall~ m \in hst.M ~ | ~ g \in m.dst \land m \notin deliveredInG$ \} }
     \ENDINDENT
        
       \vspace{1mm}
     \INDENT [g's history not informed to h so far] {\textbf{diff-hst}$(h: $ a higher group$): H$} \label{algo:funcs:diffHst}
       \STATE{let $hstTmp.M \la hst.M \setminus hst(h).M$}
       \STATE{let $hstTmp.D \la hst.D \setminus hst(h).D$}
       \STATE{let $hstTmp.lastDlvd \la hst.lastDlvd$}
       \STATE{$hst(h) \la hst$}     \COMMENT{history sent to h is updated to current history of g}
       \STATE{\textbf{return} $hstTmp$}
     \ENDINDENT

            \vspace{1mm}
   \INDENT {\textbf{depend} $(m, m': Message)$: boolean}          \label{func:depend}
       \STATE {\textbf{return} $(m',m) \in hst.D  ~ ~\lor$    }
       \STATE {\ \ \ \ \ \ \ \ \  $\exists m'' ~ | ~ (m',m'') \in hst.D ~ \land ~ $\textbf{depend}$(m, m'')$}
    \ENDINDENT
    
                      \vspace{1mm}

              \INDENT{\textbf{a-deliver} $(m: Message)$}    \label{algo:funcadeliver}
        \STATE $hst$-$add(m)$ \label{algo:funcs:addhst}
        \IF {$g = m.lca()$}
            \STATE $send$-$descendants(m, \msg)$ \label{algo:funcs:fwd}
        \ELSE
            \STATE $queues[m.lca()].dequeue()$  \label{algo:funcs:deq}
            \STATE $send$-$descendants(m, \ack)$  \label{algo:funcs:sendack}
        \ENDIF

        \IF{$ \exists [ \notif,n,deps ] \in \textit{pendNotif} ~|~ m \in deps$}  \label{algo:funcs:pendnotifs}
            \STATE $deps \leftarrow deps \setminus m$
            \IF{$deps = \varnothing$}
                \STATE {\textit{pendNotif} $\la$ \textit{pendNotif}  $\setminus  [ \notif,n,deps ] $}
                \STATE $send$-$descendants(n, \ack)$
            \ENDIF    
        \ENDIF

    \ENDINDENT

    \vspace{1mm}
    
        \INDENT{\textbf{send-descendants} $(m: Message, mType \in \{\msg, \ack\} )$} \label{algo:senddescendants}
        \STATE $send$-\textit{notifs}$(m)$ \label{algo:funcs:sendnotifs} 
                \FORALL {$descendant ~d \in m.dst$}
            \STATE $\textbf{send} ~[ mType,~m,~ $\emph{diff-hst}$(d) ] ~ \textbf{to} ~ d$ \label{algo:funcs:sendmsg}
        \ENDFOR
    \ENDINDENT
    
        \vspace{1mm}
        
    \INDENT[send \notif to groups] {\textbf{send-notifs} $(m: Message)$} \label{algo:funcs:sendnotifsBegin}
        \FORALL {$descendant ~ d ~|~ d \notin m.dst$}          
            \IF {$\exists d' \in m.dst ~ |$ \textit{d is ancestor of} $d'$
            \\ \textbf{and} $hst.containsMsgTo(d)$}
                \STATE $\textbf{send} ~ [\notif, ~ m, ~ $\emph{diff-hst}$(d)] ~ \textbf{to} ~ d$ \label{algo:funcs:sendnotifsEnd}
                \STATE {$m.$\textit{notifList.append}$(d)$}     \COMMENT{$m$ carries the notified groups}
            \ENDIF
        \ENDFOR
    \ENDINDENT
    
    
                       \vspace{1mm}
    
    \INDENT {\textbf{reprocess-queues} $()$} \label{algo:funcs:reproBegin}
        \INDENT{\textbf{do:}}
            \STATE $delivered \la false$
            \FORALL {$q \in queues$}
                \IF {$can$-$deliver(q.head())$}
                    \STATE $a$-$deliver(q.head())$ 
                    \STATE $delivered \la true$
                \ENDIF
            \ENDFOR
        \ENDINDENT
        \STATE \textbf{while} $delivered$ \label{algo:funcs:reproEnd}
    \ENDINDENT

  \vspace{1mm}

      \INDENT {\textbf{can-deliver} $(m: Message)$}          \label{func:candeliver}
        \IF{\textit{ancestors-to-ack(m)}$ ~ \nsubseteq ~ $\textit{ancestors-that-acked(m)} }    \label{canDeliverAllInfo}
            \STATE {\textbf{return} $false$}
        \ENDIF  
        \IF{$\exists ~ m' \in hst.M ~ | ~ g \in m'.dst ~ \land ~ m' \notin deliveredInG ~  \land$ \\                      \label{canDeliverNodep}
            \hspace{2.3cm} $ depend(m,m') $}
           \STATE{\textbf{return} $false$}
        \ENDIF
        \STATE \textbf{return} $true$       \label{line:canDeliver}
    \ENDINDENT
    
      \vspace{1mm}
      \INDENT {\textbf{ancestors-to-ack} $(m: Message)$: set of Groups}          \label{ancestors-to-ack}
            \STATE {\textbf{return} (ancestors of $g$ in $m.dst \setminus m.lca()) ~ ~\cup $\\ 
                        \hspace{0.8cm} $queues[m.lca()].get(m.id).$\textit{notifList}}
      \ENDINDENT

      \vspace{1mm}
      \INDENT {\textbf{ancestors-that-acked} $(m: Message)$: set of Groups}          \label{ancestors-that-acked}
            \STATE {\textbf{return} queues[m.lca()].get(m.id).acks}
      \ENDINDENT

    \end{distribalgo}
\end{algorithm}

We use set \textit{deliveredInG} to identify messages delivered in $g$ (line \ref{algo:funcs:deliverInG}).
\textit{deliveredInG} is a subset of \textit{hst.M} and is used to identify possible open dependencies in the history (line \ref{algo:funcs:hasIncompleteDeps}).
An open dependency happens when a message addressed to $g$ is included in $g$'s history but not yet delivered.
Operation \textit{diff-hst} (line \ref{algo:funcs:diffHst}) is an optimization: only the new parts of a history are sent to each descendent. 
Operation \textit{depend} (line \ref{func:depend}) computes $m$'s possible transitive dependency on $m'$ in $hst$.

When a message can be delivered (line \ref{algo:funcadeliver}), the group 
adds the message to its local history (line \ref{algo:funcs:addhst}). 
An \lca group sends the message to its descendants (line \ref{algo:funcs:fwd}),
while non-\emph{lca} groups remove the message from the ancestor's queue (line \ref{algo:funcs:deq}) 
and send the corresponding \ack messages to their descendants (line \ref{algo:funcs:sendack}).
All groups verify whether delivering this message may unblock pending notifications (line \ref{algo:funcs:pendnotifs}). 

Function \emph{send-descendants} (line \ref{algo:senddescendants}) is part of \emph{Strategies (a)} and \emph{(b)} discussed in
Section \ref{subsec:basic}.
To send \msg $m$ (or \ack $m$), the \lca (or a descendant),
first sends possible notification messages to its descendants that are not in $m.dst$.
Function \emph{send-notifs()} implements \emph{Strategy (c)}:
it searches past messages and evaluates if notifications are needed, including the notified
groups in $m$'s notification list (lines \ref{algo:funcs:sendnotifs} and \ref{algo:funcs:sendnotifsBegin}-\ref{algo:funcs:sendnotifsEnd}). 
Then, $m$ is sent to all other destinations in $m.dst$ (line \ref{algo:funcs:sendmsg}),
carrying the list of notified groups 
along with the history with information needed by each destination (\textit{diff-hst}).

Function \emph{reprocess-queues()} (lines \ref{algo:funcs:reproBegin}-\ref{algo:funcs:reproEnd}) is called upon receiving \msg and \ack messages (see Algorithm \ref{algo:events}, lines \ref{algo:events:reproc} and \ref{algo:events:repack}).

In both cases, it iterates through ancestor's queues and tries to deliver messages. 
It keeps iterating while messages can be delivered due to updated dependency information.
The delivery of messages in non-\lca groups is defined in function \emph{can-deliver(m)} (line \ref{func:candeliver}).
The first condition (line  \ref{canDeliverAllInfo}) checks whether $g$ received \ack from all needed ancestors: 
(i) all ancestors (except the \lca) in $m.dst$;
(ii) all ancestors (not in $m.dst$) \notif-ied about message $m$,
which were informed to $g$ either through \msg or \ack.   Recall that a notified group, besides sending \ack can further notify other groups.  
In Algorithm \ref{algo:events}, line \ref{algo:events:mergeNotifList}, \textit{notifList} accumulates all notified ancestors that have to \ack $m$.
The list of ancestors that have acked is kept in \textit{ancestors-that-acked} (line \ref{ancestors-that-acked}).
Having the complete information on $m$, the second condition (line \ref{canDeliverNodep}) 
ensures that any message $m'$ that precedes $m$ and is addressed to $g$ 
has already been delivered before $m$'s delivery.

\subsection{Practical considerations}
\label{sec:practic}

The protocol as described so far does not include garbage collection. 
\elia{In our \flexcast prototype, however, we prune local histories associated with each ancestor group. 
A distinguish process periodically multicast a $flush$ message to all groups. 
Once a group delivers this message, it knows that all messages that precede $flush$ can be garbage collected.
The intuition behind this mechanism is that
to deliver a message $m$ from a specific ancestor, all dependencies before $m$ must be resolved and do not need to be re-evaluated in the future.}
To further reduce communication, histories sent with messages do not enclose the ever-growing system history. 
\flexcast sends only a \emph{diff} of the history for each descendant group. 
The idea is implemented by keeping track of the last message of the local history sent to each descendant $d$ and, in subsequent messages to $d$, sending a history that contains only the newest messages added since the last communication to $d$.

\subsection{Tolerating failures}
\label{sec:ft}

\flexcast uses the same approach used in other atomic multicast protocols to tolerate failures (e.g., \cite{FastCast17}, \cite{fritzke1998amcast}, \cite{gotsman2019white}, \cite{Le2021RamCastRA}, \cite{rodrigues1998scalatom}, \cite{ByzCast}), that is, processes within a group are kept consistent using state machine replication.
This means that processes in a group can fail as long as enough processes remain operational within the group.
Consequently, groups do not fail as a whole and must remain connected (i.e., no network partition).
Tolerating the failure of a group requires additional system assumptions \cite{schiper2008solving}.

The implications of this approach on the number of correct processes per group and process communication depend of the particular consensus protocol used to implement state machine replication within a group.
For example, Paxos~\cite{Lam98} requires a majority of correct processes within each group and can tolerate message losses.






\section{Evaluation}
\label{sec:eval}

In this section, we explain the evaluation rationale, describe the environment and the benchmarks used, present the results, and summarize the main lessons learned.

\subsection{Evaluation rationale}

We compare \flexcast to a distributed atomic multicast protocol and a hierarchical atomic multicast protocol using single-process groups (i.e., no failures are tolerated) in all three protocols.
In doing so, our evaluation focuses on the inherent costs of three classes of atomic multicast protocols (see Table~\ref{tbl:amcasts}) and avoids overhead introduced by replication.
We use Skeen's protocol as distributed atomic multicast because its ordering mechanism is used by several other protocols (e.g., \cite{FastCast17}, \cite{fritzke1998amcast}, \cite{gotsman2019white}, \cite{Le2021RamCastRA}, \cite{rodrigues1998scalatom}).
Moreover, when groups contain a single process, FastCast \cite{FastCast17} and Whitebox \cite{gotsman2019white} atomic multicast protocols behave as in Skeen's protocol.
Skeen's protocol is genuine, can order messages in two communication steps, which has been shown to be optimum \cite{schiper2008inherent}, and assumes that any two groups can communicate.
We choose ByzCast as hierarchical atomic multicast protocol.
ByzCast is non-genuine and imposes a tree overlay on communication, the minimum overlay that ensures a connected system.
In single-process groups, ByzCast does not introduce any overhead particular to tolerating malicious behavior.
We implemented prototypes of all protocols in Java.

\begin{figure*}[t]
	\hspace*{-.2in}
	\includegraphics[width=.85\textwidth]{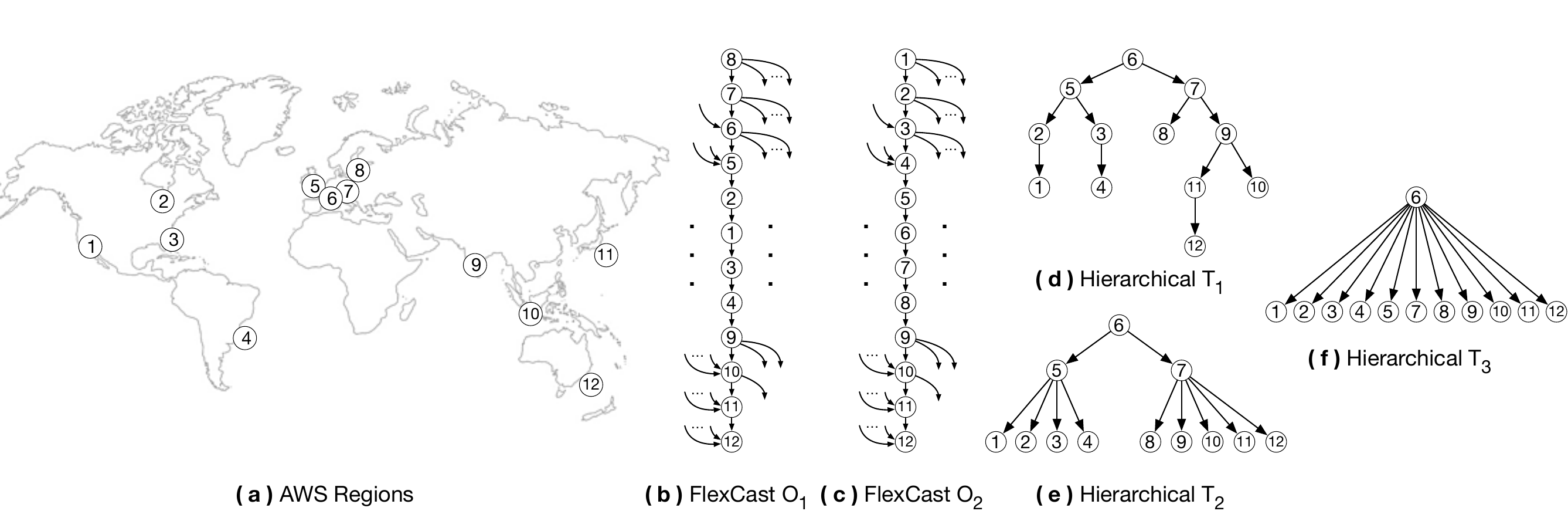}
    \caption{\small AWS regions and different overlays used in our experimental evaluation.}
    \label{fig:regions}
\end{figure*}

Our experimental evaluation aims to understand the behavior of the considered protocols in geographically distributed deployments subject to realistic workloads.
Our workload extends the well-established TPC-C benchmark to accommodate locality, a common property in geo-distributed systems.
In these settings, we seek to answer the following questions:
(i)~What is the impact of different overlays on \flexcast and hierarchical protocols?
(ii)~How quickly can a protocol order messages addressed to two or more groups?
(iii)~What is the communication overhead of hierarchical protocols?
(iv)~What is the communication cost of atomic multicast protocols?

\subsection{Environment and deployment}

\elia{The experimental setup was configured with 12 server machines and 24 client machines, connected via a 1-Gbps switched network, in CloudLab \cite{Duplyakin+:ATC19}. 
The machines are equipped with eight 64-bit ARMv8 cores at 2.4 GHz, and 64GB of RAM.
The software installed on the machines was Linux Ubuntu 20.04 (64 bits) and 64-bit Java virtual machine version 11.0.3. }
Machines communicate via TCP.

We consider an emulated wide-area network that models Amazon Web Services (AWS): 
Each group represents an AWS region and we experimented with a deployment of 12 AWS regions, as depicted in Figure \ref{fig:regions}~(a). 
The emulated latencies among regions are based on real measurements in AWS \cite{awslatencies}.
Enough client processes (to saturate our \flexcast implementation) are uniformly distributed along the 24 client machines that represent each region/group, and they send requests to the nearest group. 
Upon delivering a message, each message destination replies to the message's sender (client).


\subsection{gTPC-C Benchmark}
\label{subsec:bench}

We developed gTPC-C, a geographically distributed benchmark inspired by the well-established TPC-C benchmark \cite{tpcc}). 
We translate \tpcc warehouses into groups, deployed in one or more AWS regions, and \tpcc transactions into messages multicast to their corresponding warehouses. 

According to the \tpcc benchmark, clients can generate the following transactions (with a certain probability): new order (45\%), payment (43\%), order status (4\%), delivery (4\%), or stock level (4\%). 
The last three transactions are single-warehouse (local), resulting in a message multicast to the client's home warehouse. 
Since all multicast protocols perform the same when ordering a message multicast to a single group, in our latency measurements we only consider global transactions, which result in  messages addressed to multiple warehouses.
Consequently, this workload only contains new order and payment transactions, always involving two or more warehouses.
New order transactions can have from 5 to 15 items, where each item has a 2\% probability of being issued to a warehouse that is not the client's home warehouse, as defined by \tpcc. 

To capture locality, when choosing an additional warehouse to the client's home warehouse, the client picks the nearest warehouse to its home warehouse with a configurable high probability, the \emph{locality} rate; otherwise, the client chooses the next nearest warehouse, and so on, up to the farthest warehouse to the client's home warehouse.
Our criteria to define locality is inspired by a common wholesale supplier policy that when an item is not available in the nearest warehouse to a client (i.e., the home warehouse), it is shipped from the closest warehouse that has the item.
This locality specification implies that most messages are addressed to only two warehouses (same as in standard TPC-C), and some to three. Very few are addressed to more than three groups, therefore we do not consider these messages in our experiments.

\begin{figure*}[t]
	\begin{subfigure}{0.685\columnwidth}
		\includegraphics[width=\columnwidth]{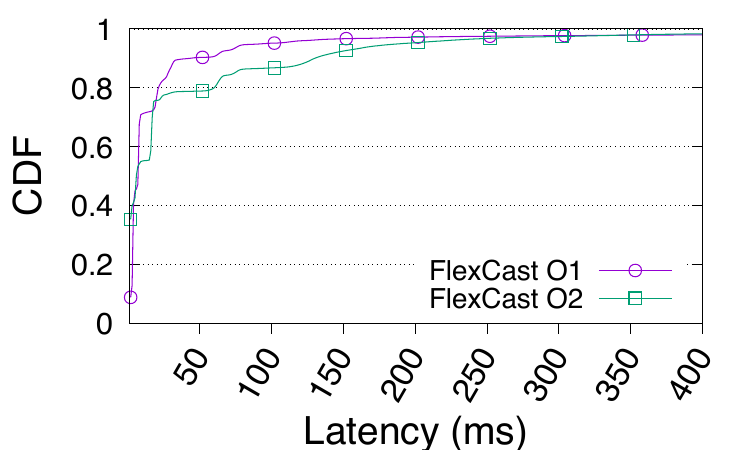}
		\caption{1st destination}
	\end{subfigure}
	\begin{subfigure}{0.685\columnwidth}
		\includegraphics[width=\columnwidth]{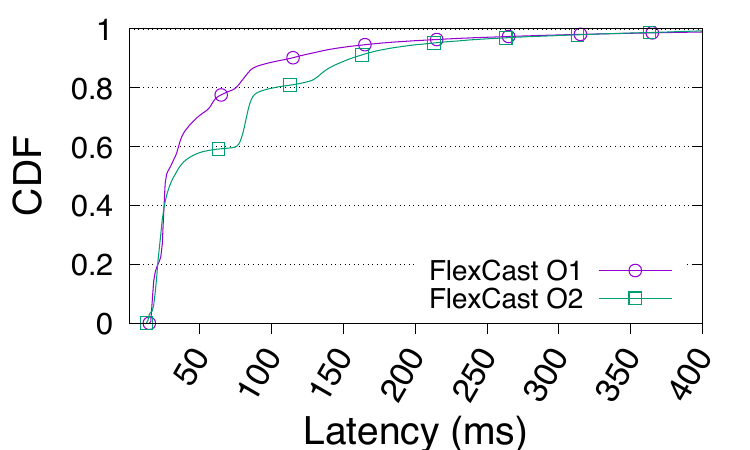}
		\caption{2nd destination}
	\end{subfigure}
	\begin{subfigure}{0.685\columnwidth}
		\includegraphics[width=\columnwidth]{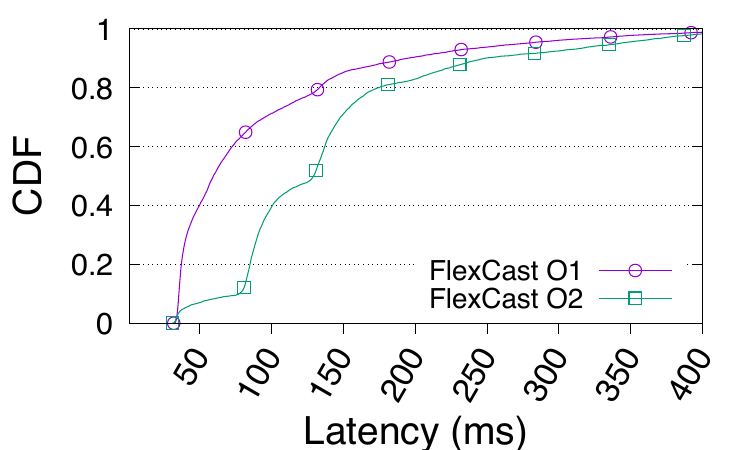}
		\caption{3rd destination}
	\end{subfigure}

	\begin{subfigure}{0.685\columnwidth}
		\includegraphics[width=\columnwidth]{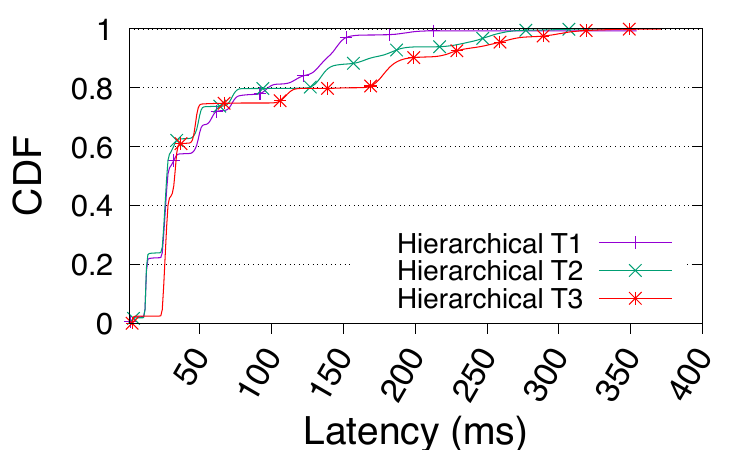}
		\caption{1st destination}
	\end{subfigure}
	\begin{subfigure}{0.685\columnwidth}
		\includegraphics[width=\columnwidth]{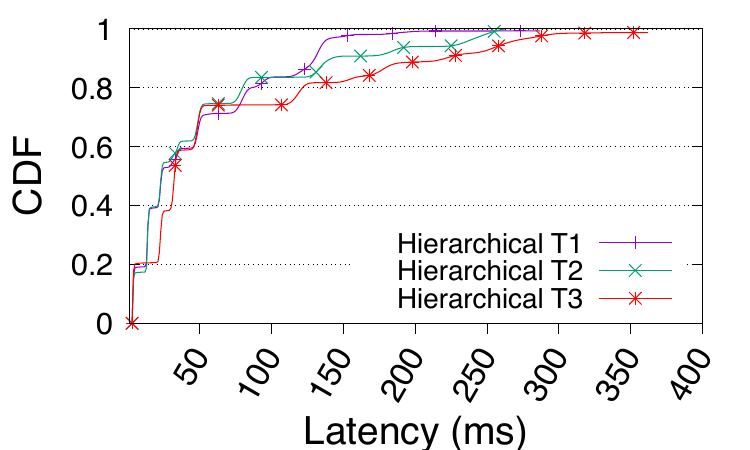}
		\caption{2nd destination}
	\end{subfigure}
	\begin{subfigure}{0.685\columnwidth}
		\includegraphics[width=\columnwidth]{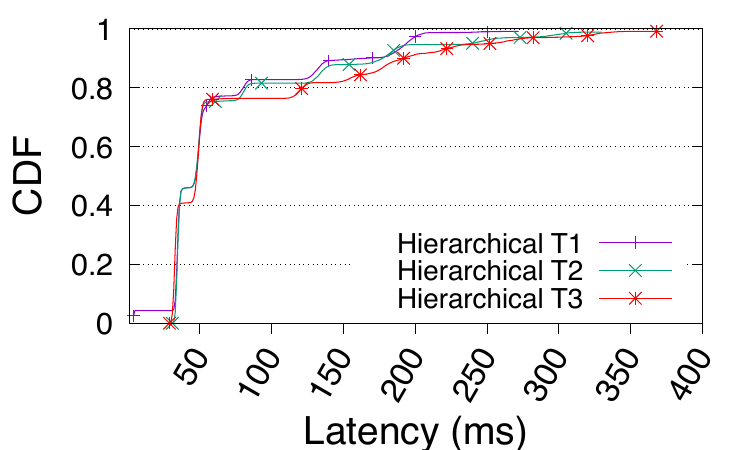}
		\caption{3rd destination}
	\end{subfigure}

	\caption{Latency per destination group when varying overlays in \flexcast and a hierarchical protocol, gTPC-C with 90\% locality.}
	\label{fig:cdftopologies}
\end{figure*}

Clients operate in a closed loop issuing one transaction at a time and are deployed in the same region as their home warehouse. 
Each experiment lasts for a period of approximately one minute, in which clients collect and store latency data. 
We discard the first and last 10\% of the data collected during the experiment to avoid possibly noisy data during warm up and end of execution.



\subsection{The effect of overlays}

In the first set of experiments, we investigate the role of overlays on \flexcast and hierarchical protocols.
We compare the latency experienced by clients of two \flexcast overlays, and three hierarchical overlays (trees), as depicted in Figure \ref{fig:regions}. 

Trees $T_1$, $T_2$ and $T_3$ contain different numbers of inner nodes.
In principle, a larger number of inner nodes provides better distribution of communication overhead among these nodes.
Trees with many inner nodes, however, may lead to additional communication steps when ordering messages.
For overlays $O_1$ and $O_2$, we initially selected a starting node (i.e., central node 8 in $O_1$ and left-most node 1 in $O_2$).
Then, the closest node to the initial one, the closest node to the second chosen node, and so on.
Since $O_1$ and $O_2$ are complete DAGs, a node is connected to all nodes that succeed it (e.g., the first node is connected to all nodes).

Figure \ref{fig:cdftopologies} and Table \ref{tbl:topologieslats} present the results. 
We report the latency per group addressed by the message.
The latency of the first (respectively, second and third) destination corresponds to the first (respectively, second and third) response the client receives
from the groups addressed by the message.
$O_1$ shows better performance than $O_2$ for all destinations.
This happens because $O_1$ better exploits locality:  higher nodes in the DAG have the lowest latencies in the geographical distribution. 
Hereafter, we evaluate \flexcast using overlay $O_1$.
 
Differently than \flexcast, whose performance is largely dependent on the overlay, a hierarchical protocol is not so sensitive to the chosen tree (but see also the discussion in Section \ref{subsec:localityeffect}), although the trees do have an impact on the performance.
$T_1$ shows slightly better performance in all destinations than $T_2$ and $T_3$. This is due to the communication overhead (further discussed in Section \ref{subsec:overhead}) of involving non-destination groups, and also the bottleneck effect of involving the tree root on $T_3$ for all messages in the system. From these results, we select $T_1$ to represent a hierarchical protocol in the rest of our evaluation.

\begin{table*}[h]
	\centering
	\small
	\begin{tabular}{|l|c|c|c|c|c|c|c|c|c|c|}
		\hline
		&  & \multicolumn{9}{c|}{Destination} \\ 
		\hline
		&  & \multicolumn{3}{c|}{1st} & \multicolumn{3}{c|}{2nd} & \multicolumn{3}{c|}{3rd} \\
		\hline
		& Overlay & 90p & 95p & 99p  & 90p & 95p & 99p  & 90p & 95p & 99p \\
		\hline
		\multirow{2}{*}{\flexcast}		
			& $O_1$ & 144.0 &  279.0 &  1403.1 &  398.0 &  829.0 &  2243.42 & 1406.0 &  2195.0 &  4542.5  \\
			& $O_2$ & 156.0 &  350.0 &  790.22  & 416.0 &  652.0 &  2006.83 & 1028.0 &  1681.5 &  3112.9 \\
		\hline
		\multirow{3}{*}{Hierarchical}		
			& $T_1$ & 229.0 &  267.0 &  311.0 & 261.0 &  288.0 &  403.0 & 307.0 &  386.0 &  408.0 \\
			& $T_2$ & 233.0 &  269.0 &  311.0 & 215.0 &  249.1 &  351.0 & 261.0 &  338.0 &  375.28 \\
			& $T_3$ & 311.0 &  398.0 &  544.0 & 381.0 &  480.0 &  622.0 & 397.0 &  531.6 &  621.0 \\
		\hline

	\end{tabular}




	\vspace{3mm}
	\caption{Latency percentiles in milliseconds for each destination group when varying the overlay in \flexcast and the tree in the hierarchical protocol, gTPC-C with 90\% locality.}
	\label{tbl:topologieslats}
\end{table*}%

\subsection{Throughput}
\elia{
In the second set of experiments, we assess the overall performance of our standard gTPC-C, including local and global messages, when deployed in a configuration with 99\% locality rate.
\begin{figure}[hb!]
	\centering
	\includegraphics[width=.68\columnwidth]{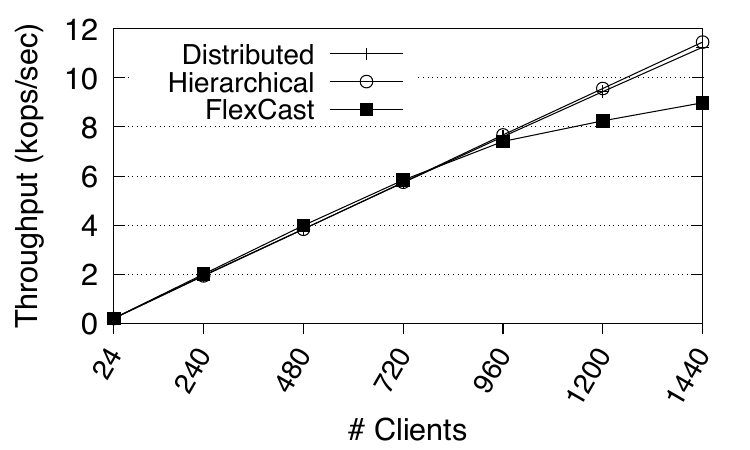}
	\caption{Throughput vs. number of clients with 99\% locality.}
	\label{fig:tp}
\end{figure}
We conduct multiple experiments while gradually increasing the number of clients and measure the total number of transactions ordered by each protocol.
Figure \ref{fig:tp} presents the results.
Although \flexcast was designed to optimize latency, it can maintain the same throughput as the other protocols up to its saturation point.
This effect can be seen by the slight bend of the throughput curve of \flexcast starting with 960 clients.
%
%
In the experiments presented next, we consider configurations with 240 clients. 
This is justified by the fact that none of the algorithms is subject to queuing effects, which would interfere with their inherent latency.
}

\subsection{Latency}
\label{subsec:localityeffect}

\begin{figure*}[t!]
	\begin{subfigure}{0.685\columnwidth}
		\includegraphics[width=\columnwidth]{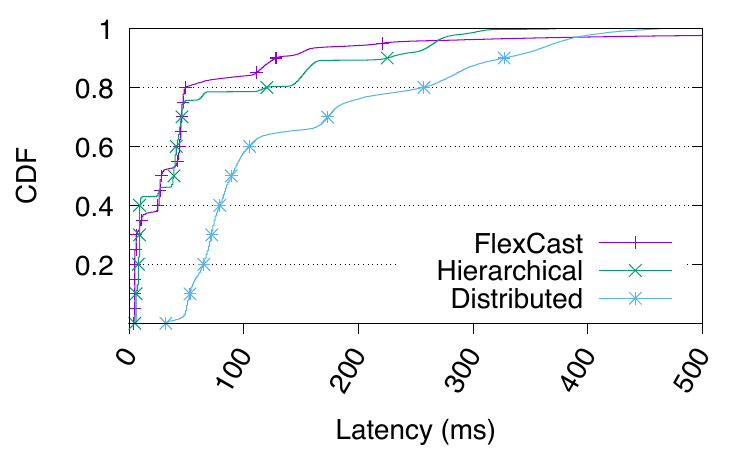}
		\caption{1st destination, 90\% Locality}
	\end{subfigure}
	\begin{subfigure}{0.685\columnwidth}
		\includegraphics[width=\columnwidth]{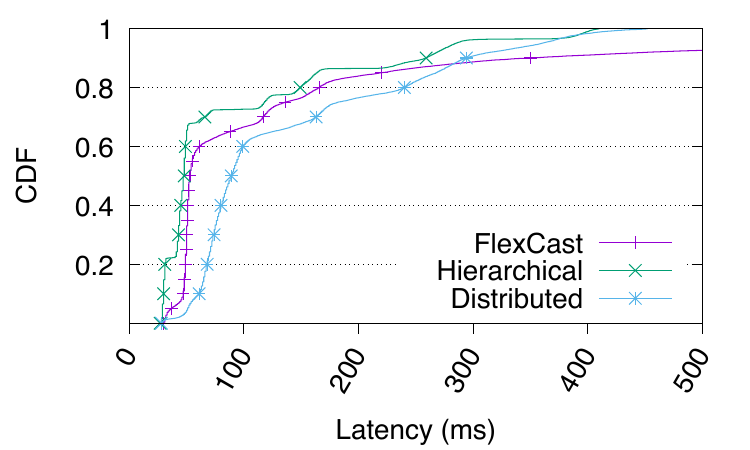}
		\caption{2nd destination, 90\% Locality}
	\end{subfigure}
	\begin{subfigure}{0.685\columnwidth}
		\includegraphics[width=\columnwidth]{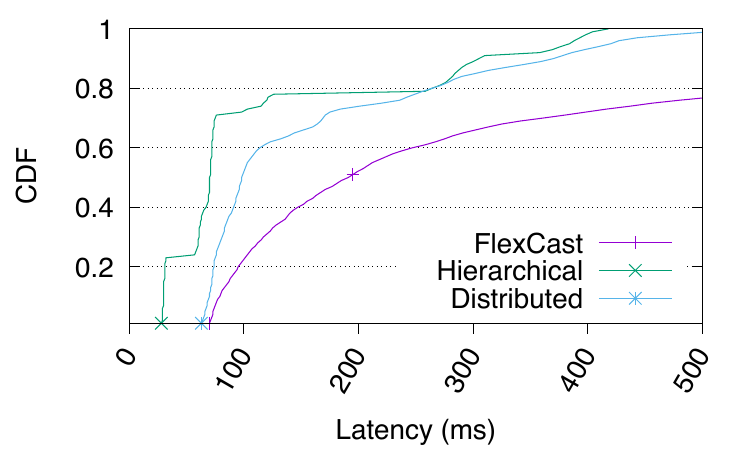}
		\caption{3rd destination, 90\% Locality}
	\end{subfigure}

	\begin{subfigure}{0.685\columnwidth}
		\includegraphics[width=\columnwidth]{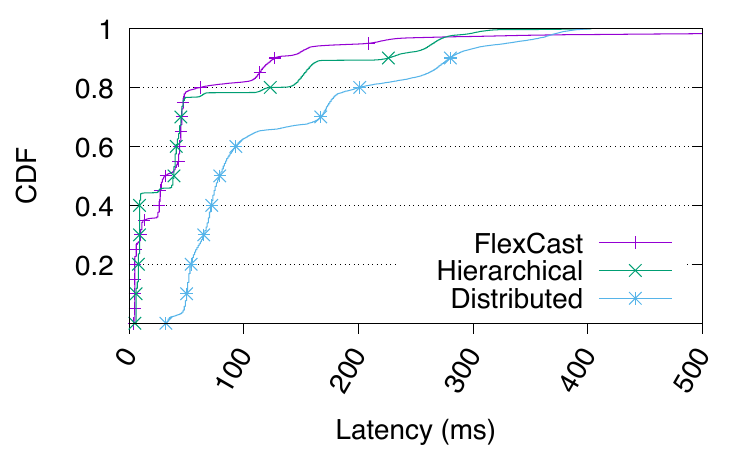}
		\caption{1st destination, 95\% Locality}
	\end{subfigure}
	\begin{subfigure}{0.685\columnwidth}
		\includegraphics[width=\columnwidth]{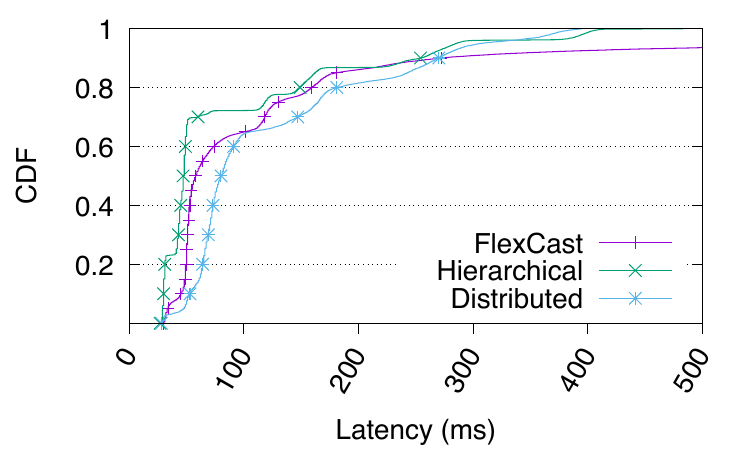}
		\caption{2nd destination, 95\% Locality}
	\end{subfigure}
	\begin{subfigure}{0.685\columnwidth}
		\includegraphics[width=\columnwidth]{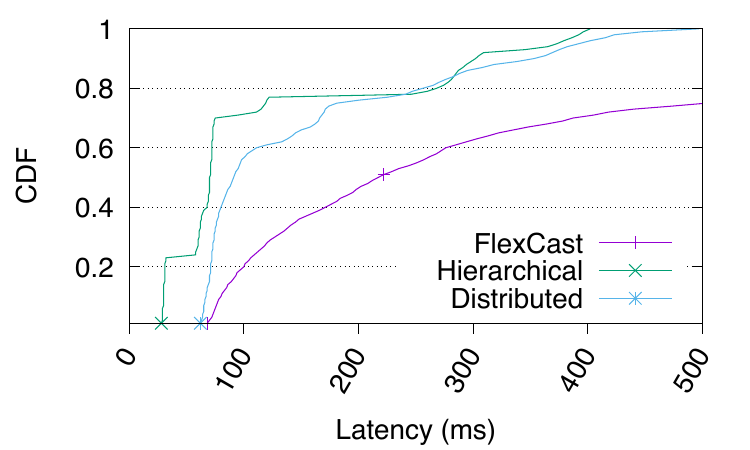}
		\caption{3rd destination, 95\% Locality}
	\end{subfigure}

	\begin{subfigure}{0.685\columnwidth}
		\includegraphics[width=\columnwidth]{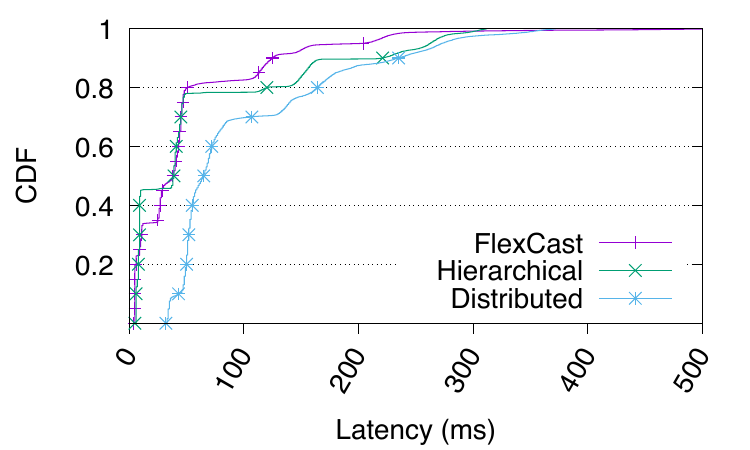}
		\caption{1st destination, 99\% Locality}
	\end{subfigure}
	\begin{subfigure}{0.685\columnwidth}
		\includegraphics[width=\columnwidth]{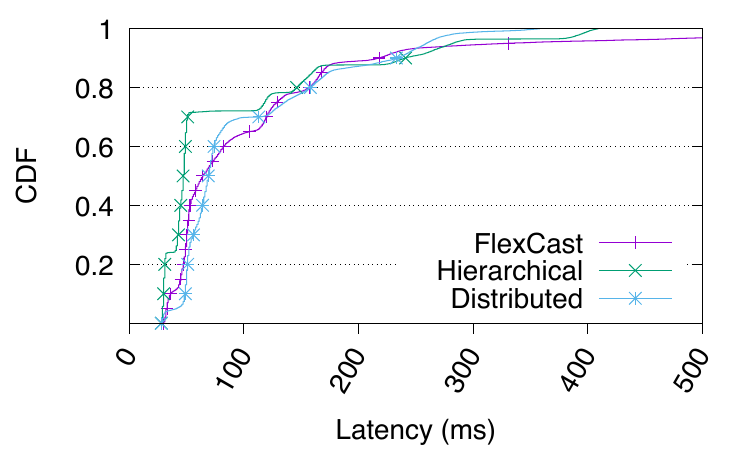}
		\caption{2nd destination, 99\% Locality}
	\end{subfigure}
	\begin{subfigure}{0.685\columnwidth}
		\includegraphics[width=\columnwidth]{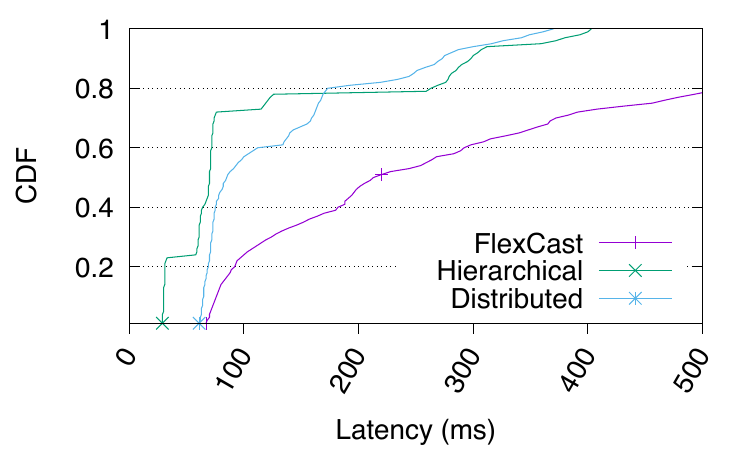}
		\caption{3rd destination, 99\% Locality}
	\end{subfigure}

	\caption{Latency per destination group when varying locality rate.}
	\label{fig:cdflocalities}
\end{figure*}

In the third set of experiments, we increase the locality rate and measure the latency experienced by the clients when receiving a response from each of the destinations of a global multicast message.
Figure \ref{fig:cdflocalities} and Table \ref {tbl:avglatency} present the results.
\flexcast outperforms both a distributed and hierarchical protocols in the latency of the first destination group for all three experimented locality rates.
We attribute this behavior to the fact that \flexcast benefits from two aspects that reduce the cost of ordering messages in the first destination in a distributed scenario: 
\emph {(i) Communication steps:} while in a distributed protocol groups addressed by a message need to exchange timestamps before a destination group can deliver a message, in \flexcast the first destination group in the DAG (i.e., the $lca$ of the message) can deliver the message as soon as it receives the message from a client; the hierarchical protocol also benefits from this aspect, however, in ByzCast, the $lca$ of a message may not be a message destination since it is not a genuine protocol.
\emph{(ii) Locality rate:} having a workload with a high locality rate increases the number of messages that \flexcast can deliver using fewer communication steps than both other protocols. This gives \flexcast an advantage since the cost for a communication step may take tens of milliseconds in geographical settings.

\begin{table*}[h]
	\centering
	\small
	\begin{tabular}{|l|c|c|c|c|c|c|c|c|c|c|c|c|}
		\hline
		&  & \multicolumn{9}{c|}{\normalsize Destination} \\ 
		\hline
		&  & \multicolumn{3}{c|}{\normalsize 1st} & \multicolumn{3}{c|}{\normalsize 2nd} & \multicolumn{3}{c|}{\normalsize 3rd} \\
		\hline
		& Locality & 90p & 95p & 99p & 90p & 95p & 99p & 90p & 95p & 99p \\
		\hline
		\multirow{3}{*}{ \flexcast} 	
			& 90\%  &  144.0 &  279.0 &  1403.1 &  398.0 &  829.0 &  2243.42 &  1406.0 &  2195.0 &  4542.5 \\
			& 95\%  &  131.0 &  217.0 &  1146.0 &  288.0 &  671.4 &  2192.64 &  1307.2 &  2231.65 &  4211.55 \\
			& 99\%  &  132.0 & 218.0  &  764.0  &  227.0 &  458.0 &  1562.09 &  1404.9 &  1975.7 &  3583.92\\
		\hline
		\multirow{3}{*}{ Hierarchical}
			& 90\% & 229.0 &  267.0 &  311.0 &  261.0 &  288.0 &  403.0 &  307.0 &  386.0 &  408.0 \\
			& 95\% & 226.0 &  265.0 &  307.0 &  255.0 &  286.0 &  403.0 &  306.0 &  381.0 &  405.0 \\
			& 99\% & 224.0 &  264.0 &  303.0 &  243.0 &  284.0 &  402.0 &  303.0 &  376.2 &  406.84 \\
		\hline
		\multirow{3}{*}{ Distributed} 	
			& 90\% & 335.0 &  377.0 &  452.0 &  299.0 &  367.0 &  444.0  &  373.0 &  423.0 &  527.7 \\
			& 95\% & 284.0 &  349.0 &  417.0 &  275.0 &  339.0 &  406.98 &  365.0 &  407.0 &  528.0 \\
			& 99\% & 241.0 &  279.0 &  370.0 &  238.0 &  263.0 &  355.0  &  309.5 &  367.0 &  415.3 \\

			


			\hline
	\end{tabular}
	\vspace{3mm}
	\caption{Latency percentiles in milliseconds for each destination when varying the locality rate for all protocols.}
	\label{tbl:avglatency}
\end{table*}%

In the second destination, \flexcast performs worse than the hierarchical protocol and outperforms the distributed protocol. 
As in the discussed above, hierarchical protocols need only one extra communication step to order a message at the second destination, while the distributed protocol, in addition to require destination groups to communicate, is also exposed to the convoy effect, which further slows down the delivery of messages \cite{gotsman2019white}.
In the third destination, \flexcast latency increases and the simplicity of a hierarchical protocol algorithm pays off. 
In both the second and third destinations, \flexcast may need extra communication steps to receive the necessary \ack messages to deliver a multicast message $m$, evaluate possible dependencies, and wait for dependencies to be solved (i.e., waiting for the delivery of previous messages ordered before $m$ in ancestor groups).
Although \flexcast performs worse than both hierarchical and distributed protocols in the third destination, messages addressed to three (or more) groups are  rare in gTPC-C, a characteristic inherited from TPC-C.

As a consequence of \flexcast's C-DAG overlay and the fact that each client in the gTPC-C benchmark is associated with the nearest warehouse, clients send most of their messages to their home warehouse and to the next nearest warehouse. 
The rate at which this phenomenon happens is regulated by the configured locality.
Therefore most messages in the workload have a disjoint destination set. 
This increases \flexcast's advantage over a distributed protocol when messages are addressed to two groups if the groups are placed consecutively in the C-DAG. 
The hierarchical protocol also benefits from locality, although as a non-genuine protocol, it introduces communication overhead, quantified in Section \ref{subsec:overhead}.
The locality rate also helps to decrease the number of auxiliary messages (i.e., \ack and \notif) needed by \flexcast to ensure consistency in the global total order, since interdependencies will be relatively fewer in such a scenario. 
Table \ref{tbl:avglatency} shows the latency percentiles (90, 95 and 99) of all destinations when varying the locality rate for all techniques. Although the hierarchical protocol shows on average a better performance when aggregating the latencies of all destinations, \flexcast is more sensitive to locality. 
\elia{
In the first destination, \flexcast's reduces 90p latency by 9\% when increasing locality from 90\% to 99\%, while the hierarchical protocol reduces by 3\%.
Despite its higher latency, the distributed protocol reduces latency by up to 29\% when increasing locality from 90\% to 99\%.
}

\subsection{The cost of exchanging histories}
\begin{figure*}[htb]
	\begin{subfigure}{0.66\columnwidth}
		\includegraphics[width=\columnwidth]{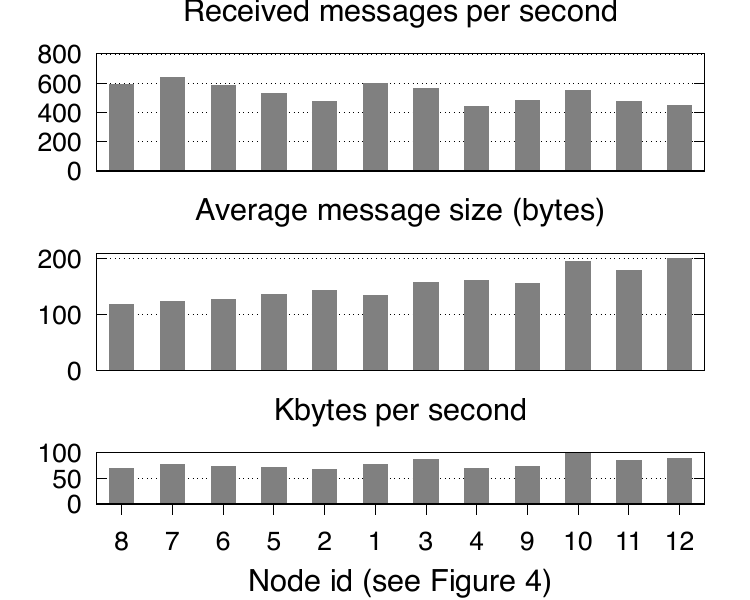}
		\caption{FlexCast}
	\end{subfigure}
	\begin{subfigure}{0.66\columnwidth}
		\includegraphics[width=\columnwidth]{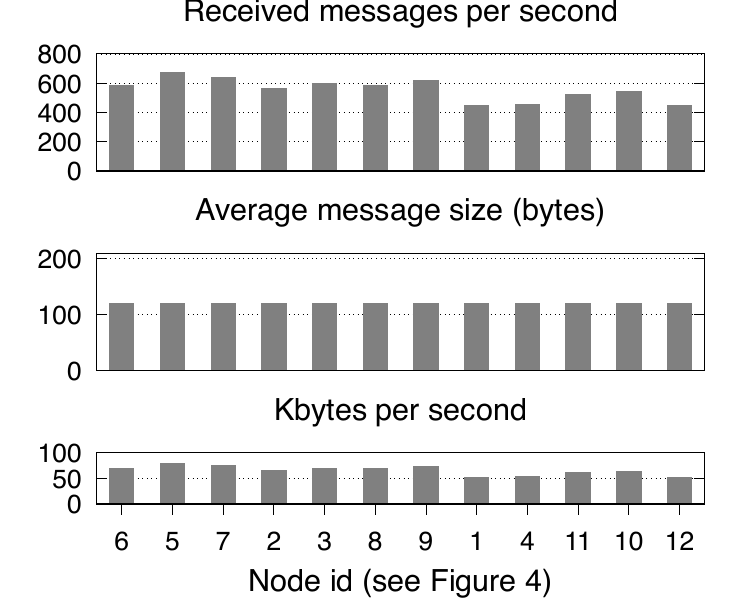}
		\caption{Hierarchical}
	\end{subfigure}
	\begin{subfigure}{0.66\columnwidth}
		\includegraphics[width=\columnwidth]{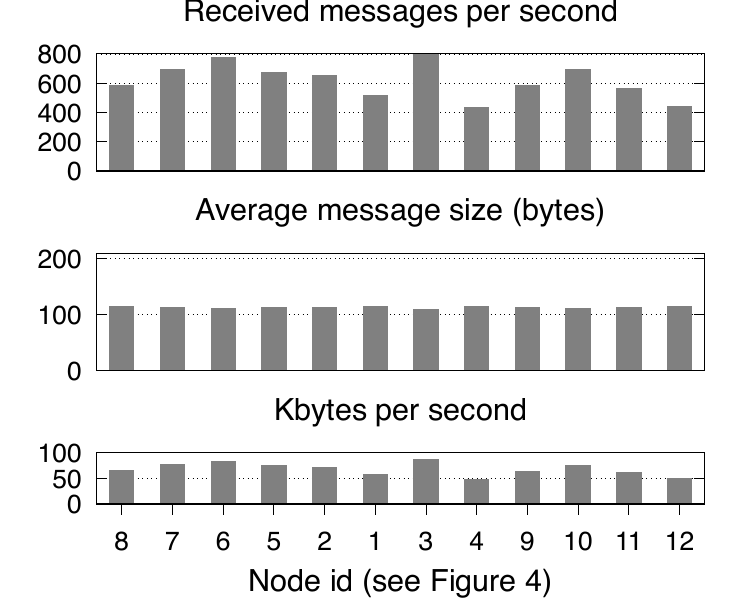}
		\caption{Distributed}
	\end{subfigure}
	\caption{The amount of information exchanged by each protocol (99\% locality, 720 clients).}
	\label{fig:msgsizes}
\end{figure*}

\elia{

In this section, we evaluate the amount of information required by each protocol to implement atomic multicast.
All protocols propagate the message payload, as defined by gTPC-C, and protocol-specific information, which in the case of \flexcast includes histories.
Figure \ref{fig:msgsizes} displays our findings. 
In each chart, the first graph (top) represents the number of messages received by each node per second.
The second graph (middle) shows the average message size per node. 
Unlike the other protocols with fixed average sizes, \flexcast shows an increase in average message size as nodes ascend the C-DAG topology (see Figure~\ref{fig:regions}).
This is due to higher nodes requiring more history data from their ancestors.
The third graph (bottom) shows the overall information exchanged by nodes per second. 

In summary, our experiments indicate that \flexcast exhibits distinctive behavior, with higher nodes in \flexcast's C-DAG exchanging a higher amount of data than lower nodes.
This results in larger messages compared to the other protocols. 
On average, a node exchanges 68.5 Kbytes per second in the distributed protocol, 66 Kbytes per second in the hierarchical protocol, and 79 Kbytes per second in \flexcast.
}

\subsection{The overhead of non-genuineness}
\label{subsec:overhead}

\begin{figure}[t]
	\centering
	\begin{subfigure}{0.65\columnwidth}
		\includegraphics[width=\columnwidth]{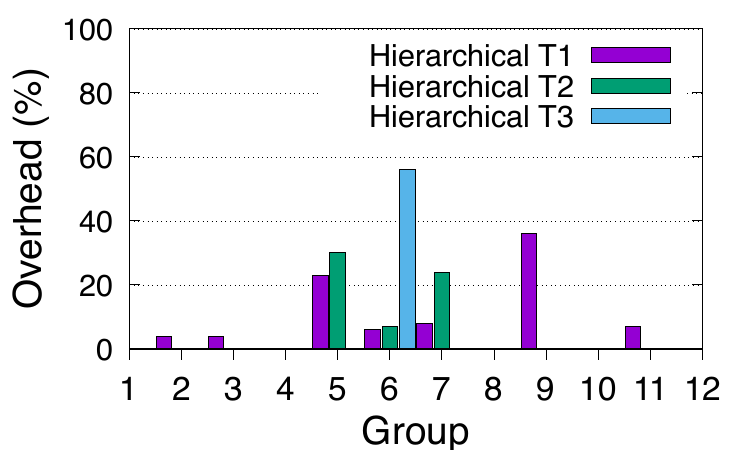}
		\caption{95\% Locality}
	\end{subfigure}

	\begin{subfigure}{0.65\columnwidth}
		\includegraphics[width=\columnwidth]{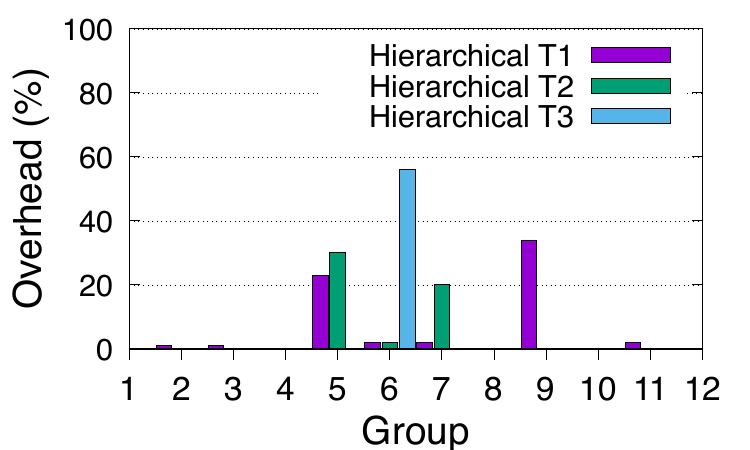}
		\caption{99\% Locality}
	\end{subfigure}

	\caption{Communication overhead of each group in hierarchical protocols with 95\% and 99\% of locality.}
	\label{fig:overheadlocalities}
\end{figure}

In this section, we investigate the communication overhead of non-genuine hierarchical protocols. 
Figures \ref{fig:overhead} and \ref{fig:overheadlocalities} present the overhead experienced per group. 
Intuitively, communication overhead captures the amount of communication involving a group due to multicast messages not addressed to the group.
We express communication overhead as a percentage and define it as 1 minus the ratio between the number of payload messages delivered by a group and the number of payload messages received by the group during an execution of the protocol.
We focus on payload messages as these are typically larger than auxiliary messages used in a protocol.

The overhead across groups depends on the tree overlay and the workload. 
But while all inner groups in a tree are potentially subject to communication overhead, leaf groups have no overhead since they are always in the destinations of messages they receive. 
Locality also plays a role in communication overhead.
A tree can benefit from locality by directly connecting groups that are near each other.
This is the motivation behind tree $T_1$:
as locality increases, $T_1$'s overhead decreases, since communication will more likely involve directly connected groups (see Table \ref{tbl:overhead}).

Tree $T_3$ has lower communication overhead than $T_1$, but this comes at the cost of penalizing group 6 (i.e., $T_3$'s root), which has to endure 56\% of overhead.
In $T_1$, groups 5 and 9 present high overhead as they are roots (lowest common ancestors) of different subtrees that represent separate geographical regions (America and Asia). 
The tree root does not have much overhead since locality is high in groups within the Europe region.
The same is observed in $T_2$, where groups 5 and 7 of disjoint subtrees present the highest overheads. 

\begin{table}[h]
	\centering
	\small
	\begin{tabular}{ |c|c|c|c| } 
		\hline
		Overlay & Locality & Mean overhead & Max \\
		\hline
		\multirow{3}{*}{\centering $T_1$} 	& 90\% & 9.16\% (11.18) & 36\% \\ 
								& 95\% & 7.33\% (11.12) & 36\% \\ 
								& 99\% & 5.41\% (11.06) & 34\% \\ 
        \hline
        \multirow{3}{*}{\centering $T_2$}  		& 90\% & 5.75\% (11.31) & 30\% \\ 
								& 95\% & 5.08\% (10.50) & 30\% \\ 
								& 99\% & 4.33\% (9.90) & 30\% \\ 
        \hline
        \multirow{3}{*}{\centering $T_3$}  		& 90\% & 4.66\% (16.16) & 56\% \\ 
								& 95\% & 4.66\% (16.16) & 56\% \\ 
								& 99\% & 4.66\% (16.16) & 56\% \\ 
		\hline
		\end{tabular}
		\vspace{3mm}
	\caption{Mean overhead, standard deviation, and maximum overhead in hierarchical trees when varying the locality rate.}
	\label{tbl:overhead}
\end{table}%

Tables \ref{tbl:topologieslats} and \ref{tbl:overhead} suggest a tradeoff: trees with the lowest latencies are subject to higher overhead on average, while trees with worse performance have lower communication overhead on average.

\subsection{Summary}
\label{subsec:summary}

We draw the following main conclusions from our experimental evaluation.

\begin{itemize}

\item \flexcast is more sensitive to the chosen overlay than the hierarchical protocol when it comes to latency.
The chosen tree, however, has an impact on the hierarchical protocol's communication overhead.

\item \flexcast consistently outperforms the distributed protocol (a genuine algorithm) in all configurations experimented.
\flexcast performs better than the hierarchical protocol in the first destination group and worse in the latency of the second and third destinations.
However, messages addressed to three (or more) groups are rare in TPC-C and gTPC-C. 
As a genuine protocol, \flexcast has no communication overhead (as defined in Section~\ref{subsec:overhead}), in contrast to a non-genuine hierarchical protocol.

\item The hierarchical protocol has a tradeoff between latency and communication overhead.
Although communication overhead is inherent to non-genuine atomic multicast protocols,
in the hierarchical protocol, trees with the best performance have the highest overhead and vice-versa.
\end{itemize}

%
%
%

\section{Conclusion}
\label{sec:conclusion}

We propose \flexcast, the first genuine overlay-based atomic multicast protocol.   
As overlay-based, it accounts for reduced connectivity in different deployment scenarios.  
As genuine, it favors geographical locality and avoids communication overhead. 
To combine both aspects, \flexcast assumes a complete DAG overlay.
Since messages may enter the overlay at different groups (nodes) of the DAG, each group takes local ordering decisions.

One interesting challenge solved by \flexcast and not yet addressed by other atomic multicast protocols is how to ensure  global acyclic order out of local ordering information from different groups.  This is achieved using a sophisticated history-based protocol.
We present \flexcast's design, its implementation, and propose a new benchmark to evaluate it: gTPC-C integrates geographical distribution and locality to the well-known TPC-C benchmark.
\flexcast shows important latency reduction in geographically distributed settings when compared to a latency-optimum genuine atomic multicast algorithm and a hierarchical protocol.

\section*{Acknowledgments}

This work was partially supported by the Swiss National Science Foundation (\# 175717), Funda\c{c}\~{a}o de Amparo \`{a} Pesquisa do Estado Do Rio Grande do Sul---FAPERGS PqG 07/21, Conselho Nacional de Desenvolvimento Cient\'{i}fico e Tecnol\'{o}gico---CNPq Universal 18/21, PUCRS-PrInt,  Coordena\c{c}\~{a}o de Aperfei\c{c}oamento de Pessoal de N\'{i}vel Superior (CAPES), Brazil, Finance Code 001, and FAPDF through EDITAL 08/2023---FAP Participa.

\bibliographystyle{acm}
\bibliography{references}

\section*{Appendix: Proof of correctness}

%
FlexCast assumes:
\begin{enumerate}
    \item \label{ass:ft-groups} that processes are organized in disjoint groups, each group being fault-tolerant;
    \item \label{ass:topology} that groups have a total order and the communication topology has directed fifo channels from each group to all higher groups.   
    \item \label{ass:ingressMsg} that when clients send a multicast message $m$ to destination groups in $m.dst$, $m$ is sent to the lowest group in $m.dst$, called the lowest common ancestor ((\emph{lca}) group.  We use $lca(m)$ to denote the lowest group
    in $m.dst$.
\end{enumerate}
Here we concentrate the discussion on the communication among groups.  Thus, saiyng that a group receives, or delivers, or sends messages means that a majority of processes in that group performs the respective action.
%
%

\vspace{2mm}
\begin{definition}  
{\it Message Order:} for any pair of messages $m \neq m'$, we say that $m < m'$ iff:
\begin{itemize}
    \item both $m$ and $m'$ are delivered at least by one same group, and $m$ is delivered before $m'$;
    \item or by transitivity:   $m < m'' \land m'' < m' \implies m < m'$.
\end{itemize}
\end{definition}

\vspace{2mm}
\begin{lemma}
    For any message $m$ atomically multicast to multiple groups, $m$ is received at all and only destination groups $d \in m.dst$.
    \label{lemma:LrcvMsg}
    \end{lemma}
    \vspace{1mm}
    {\sc Proof:} 
    By assumption \ref{ass:ingressMsg}, the client sends $m$ to $lca(m)$.   
    By assumption \ref{ass:topology}, any subset $m.dst$ of destinations is directly reached by $lca(m)$. 
    According to the algorithm, when $lca(m)$ receives $m$, it inconditionally $send$-$descendants(m)$ to all other destinations in $m.dst$ and only those.  
    As fault-tolerant groups and channels are supposed, eventually every destination group receives $m$ - and no other 
    group receives it.
    \hfill$\Box$

\vspace{2mm}
\begin{lemma}
    Let $m$ and $m'$ be messages such that $m.dst \cap m'.dst \neq \emptyset$.   
    There is a unique group that assigns a relative order to $m$ and $m'$, to be followed by all higher groups.
    \label{lemma:lcd}
\end{lemma}
    \vspace{1mm}
{\sc Proof:} 
    By assumption \ref{ass:ingressMsg}  $m$ and $m'$ ingress the overlay through their respective $lca$s
    and by lemma \ref{lemma:LrcvMsg} both  are received at their respective destinations.
    Since groups have a total order (assumption \ref{ass:topology}) 
    and $m.dst \cap m'.dst \neq \emptyset$, in the intersection there is a unique lowest group that 
    handles both $m$ and $m'$.  We call this group the lowest common destination of these messages, $lcd(m,m')$.
    Since message channels are directed towards higher groups only, the relative order of $m$ and $m'$ has 
    to be assigned at $lcd(m,m')$ and followed at higher groups, otherwise ordering could be violated.
    \hfill$\Box$
\vspace{2mm}

\begin{lemma}
    For any atomically multicast message $m$, the complete dependency information to deliver 
    $m$ is eventually received at each group in $m.dst$. 
    The {\it complete dependency information} to deliver $m$ at a group $g$ is the 
    information about any message $m'$ delivered before $m$, i.e.  $m' < m$ at each group lower than $g$. 
    \label{lemma:rcvDeps}
    \end{lemma}
    \vspace{1mm}
    {\sc Proof:} 
    By the algorithm:
    \begin{enumerate} 
        \item \label{lrcvDeps.fact-alg-1} each group $g$ keeps a history recording the order of messages it delivered and, for each message $m$ delivered, the previous messages $m'$ delivered at groups lower than $g$, such that $m'< m$;
        \item \label{rcvDeps.fact-alg-2} every message carries the history of the sending group, which enriches the history of each receiving group upon reception;
        \item \label{rcvDeps.fact-alg-3} each group $g$ in $m.dst \setminus lca(m)$ sends ACKs to higher groups in $m.dst$;
        \item \label{rcvDeps.fact-alg-4} whenever any group $g$ in $m.dst$ has previously sent messages to a group $h$ lower than others in $m.dst$, $g$ sends NOTIFY to $h$.  Each notified group $h$ reacts sending ACKs to higher groups in $m.dst$ and inductively behaves as $g$ to NOTIFY further groups.  
        Since groups have a total order, this induction finishes.
    \end{enumerate} 

    From Lemma \ref{lemma:LrcvMsg} and facts above, it follows that each group in $m.dst$ is provided with the history of each lower group 
that may be involved in messages ordered before $m$.
\hfill$\Box$
\vspace{2mm}

\begin{lemma}
    For any atomically multicast message $m$, any destination group in $m.dst$ 
    knows when the complete dependency information has been received.
    \label{lemma:completeDeps}
    \end{lemma}
    \vspace{1mm}
    {\sc Proof:} 
    By Lemma \ref{lemma:LrcvMsg} each group in $m.dst$ receievs $m$, by the algorithm
    it knows which are the lower groups in $m.dst$ and awaits for their respective ACKs.  
    Each ACK informs also if the sending group has notified
    other groups, from which by the algorithm further ACKs are awaited 
    (see Lemma \ref{lemma:rcvDeps}, facts \ref{rcvDeps.fact-alg-3} and \ref{rcvDeps.fact-alg-4}).   
    Thus, from the messages received, any destination of $m$ is able to detect if it has received 
    $ACK$s from all groups with messages ordered before $m$.
    \hfill$\Box$
    \vspace{2mm}

\begin{proposition}
(\textit{\flexcast is Genuine})
A multicast protocol is said genuine if, in a run $R$, only the message sender and destinations should communicate 
to propagate and order a multicast message.
\end{proposition}

\vspace{2mm}
{\sc Proof:} 
From the algorithm, when $m$ is multicast, there are three kinds of messages 
possible in the overlay: \msg, \ack and \notif.
\msg and \ack messages are exchanged exclusively among groups in $m.dst$, 
i.e. it's destinations.
A \notif message can only be sent from a group $g \in m.dst$ to $h$ if there exists a 
previous message $m'$ in run $R$ and $\{ g, h \} \in m'.dst$.   It follows thus that only
destinations of messages in $R$ communicate propagate and order 
their messages.
\hfill$\Box$
\vspace{2mm}

\begin{proposition}
    (\textit{Validity and Agreement}) 
\end{proposition}
{\sc Proof:} 
Due to assumption \ref{ass:ft-groups}, Lemmas \ref{lemma:LrcvMsg}, \ref{lemma:rcvDeps} and \ref{lemma:completeDeps},
and by the algorithm, we have that all groups in $m.dst$ eventually have $m$ and are able to pass the evaluation of the first condition of {\bf can-deliver(m)}.   
It remains to check if there is any message $m'$ that should be deliverd before $m$.
If no $m'$ exists, then the group can deliver $m$.
If there exists such $m'$ it has to be first delivered.  
Assuming acyclic order, which is further discussed, the arguments above
and by induction on message dependecies, there will allways be
a message with no pending dependencies to deliver that will then 
enable further ones to be delivered, such that $m$ can be delivered.
Therefore, validity holds.
By the same arguments, agreement holds.
\hfill$\Box$
\vspace{2mm}

\begin{proposition}
    (\textit{Integrity}) 
\end{proposition}
{\sc Proof:} 
By Lemma \ref{lemma:LrcvMsg} a multicast message $m$ reaches all and only its destination groups.
Any other possible message (Acknowledgements or Notifications) do not convey messages
to be delivered.   So, a group $g$ delivers $m$ only if $g \in m.dst$ and $m$ has been multicast first.
\hfill$\Box$
\vspace{2mm}

\begin{proposition}
    (\textit{Prefix Order}) 
\end{proposition}
\vspace{2mm}
{\sc Proof:} 
    From Lemma \ref{lemma:lcd} there is a unique group, $lcd(m,m')$, that assigns the
    relative order among $m$ and $m'$.
    From Lemmas \ref{lemma:rcvDeps} and \ref{lemma:completeDeps} any further group in 
    $h \in m.\mathit{dst} \cap m'.\mathit{dst}$ receives and preserves 
    the order assigned by $lcd(m,m')$.   Thus prefix order holds.
\hfill$\Box$
\vspace{2mm}

\begin{proposition}
    (\textit{Acyclic Order}) 
\end{proposition}
\vspace{2mm}

To argue that \flexcast ensures acyclic order we use a contradiction.
Assume cycle $C$ exists: $m_1 < m_2 < ... < m_k < m_1$.
Let $C$ be such that $m_k < m_1$ happens at group $h$ (i.e., $h$ delivers $m_k$ and then $m_1$), where  $h$ is the highest group in the overlay.
This is possible because the overlay induces a total order on groups.

\begin{figure}[ht]
	\centering
    \includegraphics[width=.9\columnwidth]{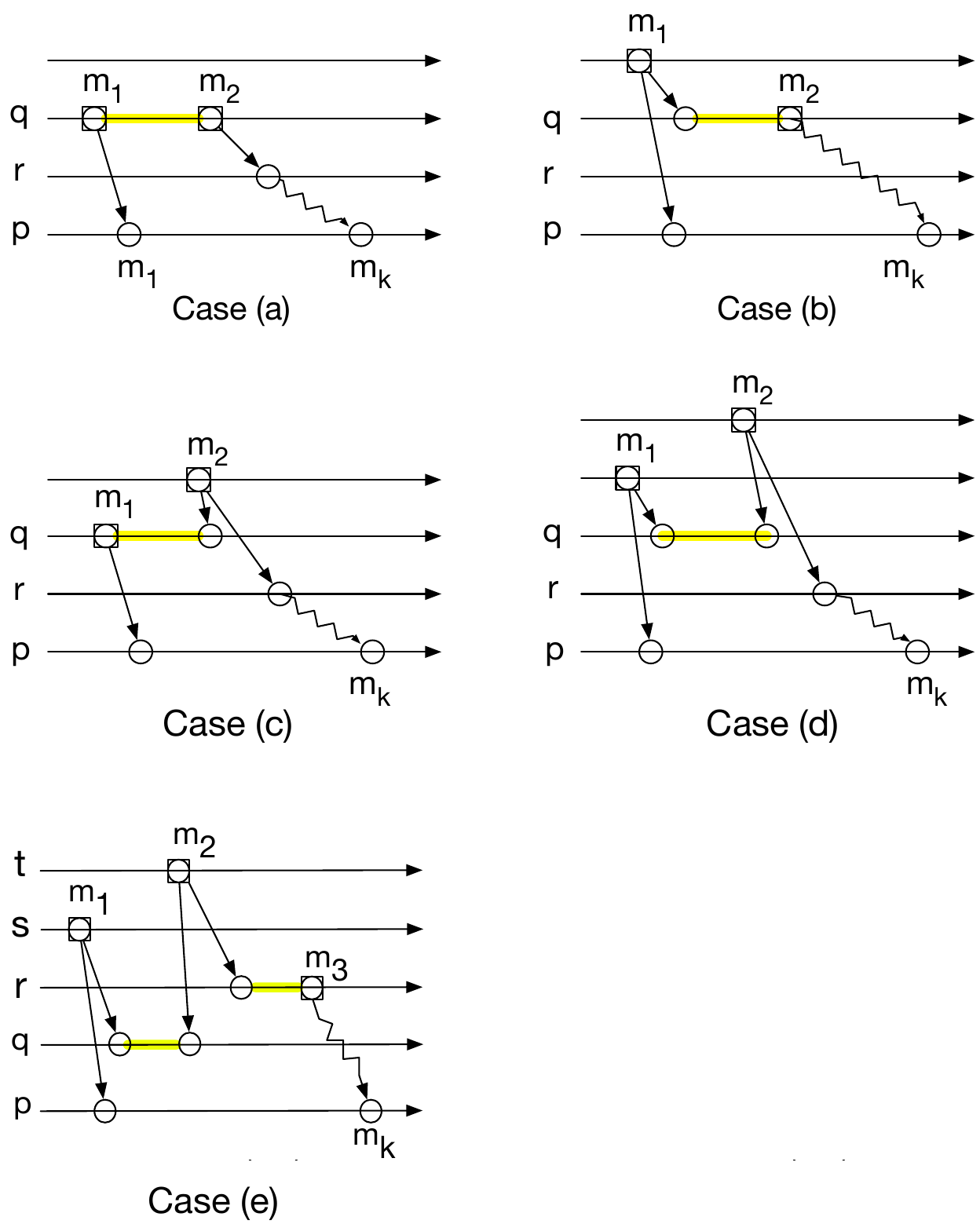}
    \caption{Causal paths}
    \label{fig:causalpath2}
\end{figure}

Let $q$ be the $lcd$ group that delivers messages $m_1$ and $m_2$.   We consider all $lca$ combinatios for $m_1$ and $m_2$ (in Figure \ref{fig:causalpath2}, cases a, b, c and d).
We claim that there is a causal path $P$ from the delivery of $m_2$ at $q$ to the reception of message $m_k$ at process $p$.

Since processes deliver messages following their causal dependencies, showing that causal path $P$ exists means that before $p$ delivers $m_k$, it knows that $m_1$ precedes $m_k$, which leads to a contradiction since $p$ will not deliver $m_k$ before delivering $m_1$.

\vspace{2mm}

The proof of the claim is by induction on the size of cycle $C$.

\vspace{1mm}

\emph{Base step ($k=2$):} This case corresponds to the four patterns involving messages $m_1$ and $m_2$ (see Figure \ref{fig:causalpath2}), having $r=p$. 
For patterns (a) and (b), the claim follows directly.
For patterns (c) and (d): Since $m_2$ is addressed to $q$ and $p$, and $p$ is below $q$ in the overlay, upon delivering $m_2$, accoording to the algorithm, $q$ sends an ACK message to $p$ (with all $q$'s dependencies) and thus there is a causal path.

\vspace{1mm}

\emph{Inductive step:}
Assume there is a causal path between $m_2 < m_3 < ... < m_k$. 
We show that there is a causal message path from $m_1$ to $m_k$, 
where $q$ delivers messages $m_1$ and $m_2$, and $r$ is one of the destinations of $m_2$ (together with $q$ and possibly other processes). 

There are five possibilities for how $q$ creates a dependency between $m_1$ and $m_2$, and where $r$ is placed with respect to $q$ in the communication overlay (see Figure \ref{fig:causalpath2}).
\begin{itemize}
\item Cases (a) and (b). In these cases, $r$ is necessarily below $q$ in the overlay, since $q$ multicasts $m_2$ and otherwise $r$ would not be a destination of $m_2$. In these cases, $m_2$ multicast by $q$ to $r$ creates a causal path from $m_1$ to $m_2$ at $r$. 
From the induction hypothesis, this leads to a causal path until $m_k$.

\item Cases (c) and (d). In these cases, we consider that $r$ is below $q$ in the overlay. 
Since both $q$ and $r$ are destinations of $m_2$ and $r$ is below $q$, from the algorithm, $q$ sends an ACK message to $r$ and $r$ waits for the ACK message before delivering $m_2$. 
This creates a causal path between the delivery of $m_1$ and $m_2$ at $q$ and the delivery of $m_2$ at $r$. 
From the induction hypothesis, it follows that there is a causal path all the way to the delivery of $m_k$ at $p$.

\item Case (e). $r$  is positioned above $q$ in the communication overlay.  
Since there is a causal path $P$ between the delivery of $m_2$ at $r$ and the receive of $m_k$ at $p$, it is the case that $r$ sent a message in $P$, say $m_3$.
Regarding  the generation of $m_3$, it could also be that $r = t$.
Regarding the generation of $m_1$, it could be that  $s=q$.

Since $r$ knows that it was involved in $m_2$ with $q$, below $r$ in the overlay, $r$ sends a NOTIFY message to $q$, and as a response, $q$ sends an ACK message in path $P$ to groups in $m_3.dst$ below $q$ (completing the information that $m_1$ is in the past of $m_3$).   Since groups can only deliver $m_3$ once these ACKs arrived, further messages after $m_3$ build a path $P$ to $m_k$ in $p$ starting from $m_2$ in $r$.
From the induction hypothesis that there is a path from $m_1$ to $m_k$.

\end{itemize}


\end{document}